\documentclass[final, 5p, times, twocolumn]{elsarticle}

\usepackage[T1]{fontenc} 
\usepackage{float}
\usepackage{array,multirow}
\usepackage{comment}
\usepackage{amsmath}
\usepackage{mathtools}

\usepackage{hyperref}
\hypersetup{
    colorlinks=true,
    linkcolor=blue,
    filecolor=magenta,      
    urlcolor=cyan,
}
 
\usepackage{afterpage}
\usepackage{placeins}
\usepackage{graphicx}
\graphicspath{ {./images/} }
\usepackage{caption}
\usepackage{subcaption}

\usepackage{gensymb}
\usepackage{graphicx}
\usepackage[utf8]{inputenc}
\usepackage{url}
\usepackage{bm}
\usepackage{booktabs}
\usepackage{lineno}
\usepackage{ulem}

\journal{Astroparticle Physics}

\begin{document}
\begin{frontmatter}
\title{Simulation of ionizing radiation in cell phone camera image sensors}

\author[a, b]{Runze Li\corref{correspondingauthor}}
\author[a,b]{Alex Pizzuto\corref{correspondingauthor}} 
\author[a,b]{Justin Vandenbroucke\corref{correspondingauthor}}
\author[a, b]{Brent Mode} 

\cortext[correspondingauthor]{Corresponding authors: \url{runze.li@icecube.wisc.edu}, \url{pizzuto@icecube.wisc.edu}, \url{justin.vandenbroucke@wisc.edu}, }

\address[a]{Department of Physics, University of Wisconsin-Madison, Madison, WI 53706, USA}
\address[b]{Wisconsin IceCube Particle Astrophysics Center, Madison, WI, 53703, USA}


\begin{abstract}
The Distributed Electronic Cosmic-ray Observatory (DECO) is a cell phone app that uses a cell phone camera image sensor to detect cosmic-ray particles and particles from radioactive decay. Images recorded by DECO are classified by a convolutional neural network (CNN) according to their morphology. In this project, we develop a GEANT4-derived simulation of particle interactions inside the CMOS sensor using the Allpix$^2$ modular framework. We simulate muons, electrons, and photons with energy range 10 keV to 100 GeV, and their deposited energy agrees well with expectations. Simulated events are recorded and processed in a similar way as data images taken by DECO, and the result shows both similar image morphology with data events and good quantitative data-Monte Carlo agreement.

\end{abstract}


\end{frontmatter}

\section{Introduction}
\label{sec:intro}
Commercial and industrial demand for silicon-based sensors has led to an exponential growth in instrumented area of silicon. Although much of the innovation in silicon-based technologies can be linked to personal use applications, scientific pursuits such as medical physics, high energy physics, and multi-messenger astrophysics have benefited vastly from the recent improvements. Many of these advancements benefit both academic and industrial pursuits, however differences in technical requirements often limit the possible interdisciplinary uses of a particular device.

The Distributed Electronic Cosmic-ray Observatory (DECO) \cite{DECO} seeks to capitalize on the vast area of instrumented silicon in commercial devices for academic and educational purposes by using camera image sensors to detect ionizing radiation produced by both cosmic rays and radioactive decay. As a cell phone application, DECO records a camera image once every 1-2 seconds with the camera lens covered to prevent contamination from background light. When ionizing radiation interacts with the CMOS chip, according to the mechanism described in section \ref{sec:APS}, electron-hole pairs are generated.  The charge is drifted to the silicon surface and recorded as an image. A convolutional neural network (CNN) \cite{Winter:2018enx} is used to classify the images recorded by DECO into three morphological classes: tracks, worms, and spots. These definitions follow the convention in \cite{define_morphology}. Tracks are long straight clusters of pixels, usually produced by minimum ionizing particles with energies at the GeV scale. At sea level, these particles are mainly muons from cosmic rays. Worms are curved clusters of pixels, usually due to low-energy electrons which undergo multiple Coulomb scatterings. Spots are small circular clusters of pixels, and they can be produced by alpha particles, Compton scattering of photons, or cosmic rays incident normal to the sensor. To date, DECO data from all seven continents have led to the identification and classification of various forms of ionizing radiation \cite{Winter:2018enx} as well as measurements of detector geometry from the known angular distribution of cosmic-ray muons at sea level \cite{Vandenbroucke:2015kkk}.


Although the CNN in DECO can classify events based on their morphology, training the CNN relies on human labeled events, so the size of available training data is limited \cite{Winter:2018enx}. Also, since humans can only label events based on their morphology, the CNN cannot be used to classify the particle type or energy of images. One possible solution to these problems is to use simulated events to train the CNN instead of human-labeled data events recorded by DECO. Since simulated events include the true particle type and energy, a CNN may be able to classify these labels as well. In this work, we present a simulation of a cell phone CMOS sensor including beam production, interactions within the silicon pixels, electron-hole transport, signal digitization, and image processing.  We use the Allpix$^2$ modular simulation framework built on top of GEANT4. With various simulation parameters, including pixel size, electric field strength, and image processing algorithms, a Monte Carlo simulation of particles produced either by cosmic-ray interactions or radioactive decay (muons, electrons, and photons) is used to produce simulated DECO images. The results are fed into the same CNN for image morphology classification, and simulated events are compared with data events. In Section~\ref{sec:APS} we review the relevant physics of silicon pixels, and then discuss our specific simulated detector in Section~\ref{sec:allpix}. We then present the processing of raw images in Section~\ref{sec:image_process}; the Monte Carlo framework is described in Section~\ref{sec:MC_sim}. The results of this simulation are shown in Section~\ref{sec:results}, which includes a study of deposited energy (Section~\ref{sec:energy_rec}), image examples from simulation and data events (Section~\ref{sec:single_comp}), classification distribution (section~\ref{sec:prob_dist}), and quantitative data-Monte Carlo comparisons (Section \ref{sec:dist_comp}).

\section{Simulation Configuration}

\subsection{Active Pixel Sensors}
\label{sec:APS}
Nearly all semiconductor devices rely on using junctions of n- and p-type silicon to form diodes. Although bulk n- and p-type silicon are electrically neutral, n-type silicon has an excess of electrons in the conduction band while p-type silicon has an excess of holes in the valence band. When these two materials are joined together, thermal fluctuations lead to a diffusion of the excess charges to the region of opposite type. This creates an excess of negative charge in the p-type region and an excess of positive charge in the n-type region, creating an electric field between the two. The extent of this electric field determines the size of the depletion region.

If ionizing radiation creates electron-hole pairs in the depletion region, then the electric field drifts the charges in opposite directions. These charges can be gathered over a fixed integration time, and then digitized and read out as a signal. Although most active pixel sensors used for astrophysical purposes use charge-coupled devices (CCDs), most cell phone image sensors are complementary metal-oxide sensors (CMOS). A key difference is that CCDs, after an exposure, transport the charges deposited on each pixel to a common set of readout electronics. The charges are physically transported, and then read out systematically on the corner of a chip with many pixels \cite{CCD_CMOS_comp}. CMOS detectors, on the other hand, have transistors placed directly in contact with each pixel, allowing one to read out pixels individually. This distinction is not crucial for this simulation, but the underlying digitization of triggering a pixel based on charge deposited on each pixel is similar in logic to the CMOS implementation, which is what we would like to emulate. Additionally, in this framework, thresholds can be smeared out to reflect imperfections in read-out electronics from pixel to pixel. This was not implemented in this analysis, but would be an interesting investigation for future study, and may more accurately reflect physical detectors with imperfections.

\subsection{GEANT4 and Allpix$^2$ Simulation}
\label{sec:allpix}

While the expected number of electron-hole pairs created by minimum ionizing particles can be calculated with the Bethe-Bloch formula and the charge creation energy of the sensor material, the underlying particle interactions are stochastic in nature, and are most easily integrated using Monte Carlo techniques. This is accomplished using Allpix$^2$, a C++ open-source modular framework for the simulation of silicon pixel detectors \cite{Spannagel:2018usc} \footnote{code is available at \cite{AllpixCode}}. Allpix$^2$ interfaces with GEANT4 \cite{Agostinelli:2002hh}, a simulation toolkit which handles the particle interactions in matter, and requires ROOT \cite{Brun:2002ifj}. Modules can be selected to choose the type of silicon pixel, the dimensions of the pixel array, electric fields, charged particle transport, and digitization. In this project, a pixel array similar to the the CMOS in an HTC Wildfire cell phone is simulated, with details of physical parameters listed in Table \ref{tab:sim_parameter}. During previous research for DECO \cite{Vandenbroucke:2015kkk}, the physical parameters in an HTC Wildfire were examined. There is a database of over 1000 images taken with the HTC Wildfire, which provides a large enough experimental data set to compare against this simulation.

\begin{table*}[h!]
    \caption{Parameters of the simulated device. Many parameters are chosen based on the analysis reported in \cite{Vandenbroucke:2015kkk}. Some values are estimated as much of the information on the readout electronics in cell phone models is proprietary.}
    \label{tab:sim_parameter}
    \centering
    \begin{tabular}{l | c } \hline
     Parameter & Value  \\ \hline \hline
     Number of Pixels & 2592 $\times$ 1944 (5,038,848) \\
     Pixel Size & 0.9$\mu$m $\times$0.9 $\mu$m \\
     Depletion Thickness & 26.3 $\mu$m \\
     Temperature & 293 K \\
     Surrounding Material & Vacuum \\
     Electric Field Model & Linear \\
     Bias Voltage & -25 V \\
     Integration Time & 50 ns \\
     Max Step Size & 1 $\mu$m \\
     Charge Creation Energy & 3.62 eV \\
     Charge Threshold as Hit & 1 e \\
     QDC Resolution & 8 bits \\
     QDC Slope & 0.5 e \\
     QDC Smearing & 0 e
    \end{tabular}
\end{table*}

In this simulation, the detector is modeled as a silicon monolithic CMOS in vacuum. This simplified model ignores the surrounding materials of the cell phone.  In real experimental data, the surrounding materials may act as a source of radioactivity and may absorb or scatter particles incident from outside the cell phone.  In the simulation, particles are initialized in the center of the detector top surface, with certain zenith angle, azimuth angle, and energy. Physical imperfections and electronics noise are not simulated directly but added later during image processing in order to match the noise level in a real device. For electric field configuration inside the CMOS, a simplified linear electric field is used due to insufficient knowledge about the commercial CMOS configuration, and the bias voltage is adjusted based on results in \ref{sec:dist_comp}. The integration time is the amount of time in which charges reaching the surface will be counted, and 50 ns guarantees that most of the charges created will be collected. The charge creation energy of 3.62 eV is the energy required to create an electron-hole pair in silicon CMOS, and the charge threshold parameters is set to $1$ so that all charges that reach the surface are counted. The QDC converter converts the charge of each pixel to a digital value (``luminance''). The QDC slope of $0.5e$ means that the amount of charge is scaled by a factor of 2 to calculate the luminance with 8 bits of QDC resolution.  QDC saturation is included: after scaling, any luminance larger than 255 is set to 255. The QDC smearing controls the noise in the QDC, and here a perfect conversion is assumed.

\begin{figure}[h!]
\begin{subfigure}{0.5\textwidth}
  \centering
  \includegraphics[scale = 0.5]{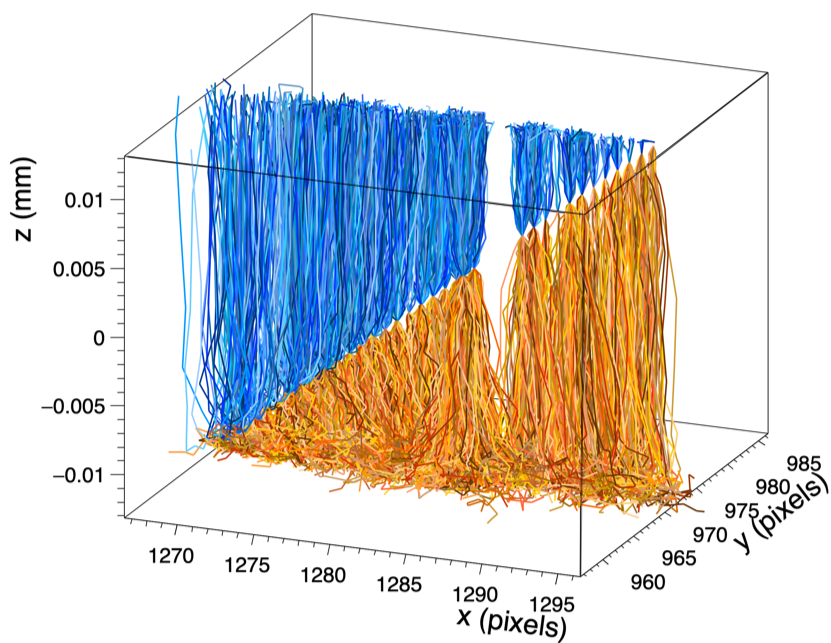}
  \caption{}
\end{subfigure}
\begin{subfigure}{0.5\textwidth}
  \centering
  \includegraphics[scale = 0.21]{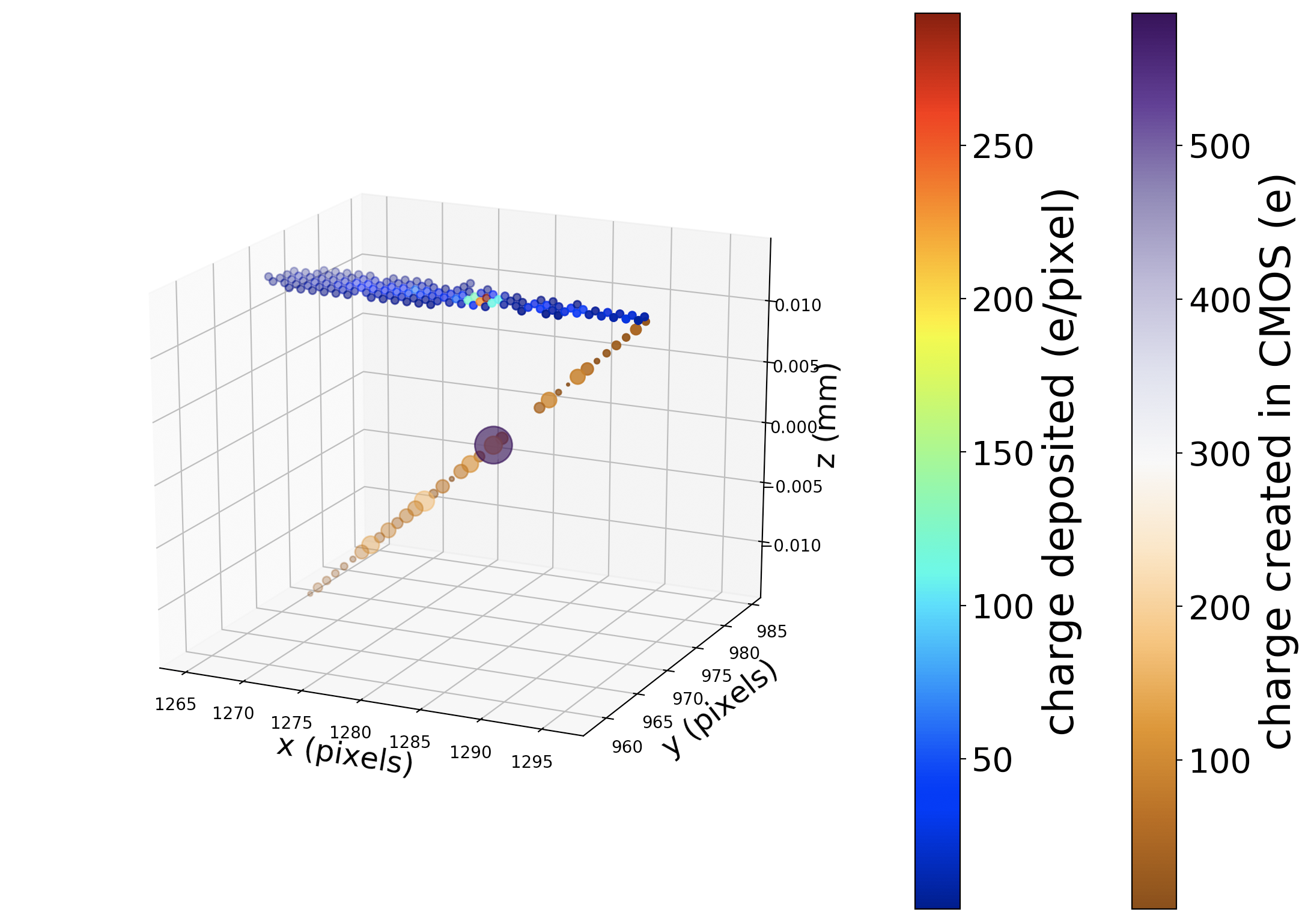}
  \caption{}
\end{subfigure}

\caption{Simulation of a 4 GeV $\mu^+$ interacting with the camera image sensor. Figures (a) and (b) are based on the same event. In (a) the charge propagation inside the CMOS sensor is shown.  The ionizing radiation generates electrons (blue) and holes (orange) that drift to opposite surfaces due to the bias voltage. In (b) the electron creation and deposition inside the CMOS sensor is plotted, with the size and color of each sphere along the muon track indicating the number of electrons created and the color map at the top surface representing the number of electrons integrated in each image pixel.  The stochastic nature of the ionization loss is evident.}
\label{fig:muons_detector}
\end{figure}

As an example, a simulated 4~GeV muon (representative of a cosmic-ray muon at sea level) interacting with the detector are shown in Figure~\ref{fig:muons_detector}. As shown in figure \ref{fig:muons_detector} (a), when a 4 GeV muon passes through the CMOS model, electrons and holes are generated by ionizing radiation, and the linear electric field propagates them to opposite surfaces. On the top surface, deposited electrons are binned into pixels as shown in Figure~\ref{fig:muons_detector} (b), and a QDC with parameters listed in table \ref{tab:sim_parameter} converts the charge of each pixel to luminance. After the QDC, the luminance of the raw images is an array of size $2592 \times 1944 \times 1$, and this raw image is used for further processing.

\subsection{Image Processing}
\label{sec:image_process}

\begin{figure}[h]
\centering
  \includegraphics[width=0.35\textwidth]{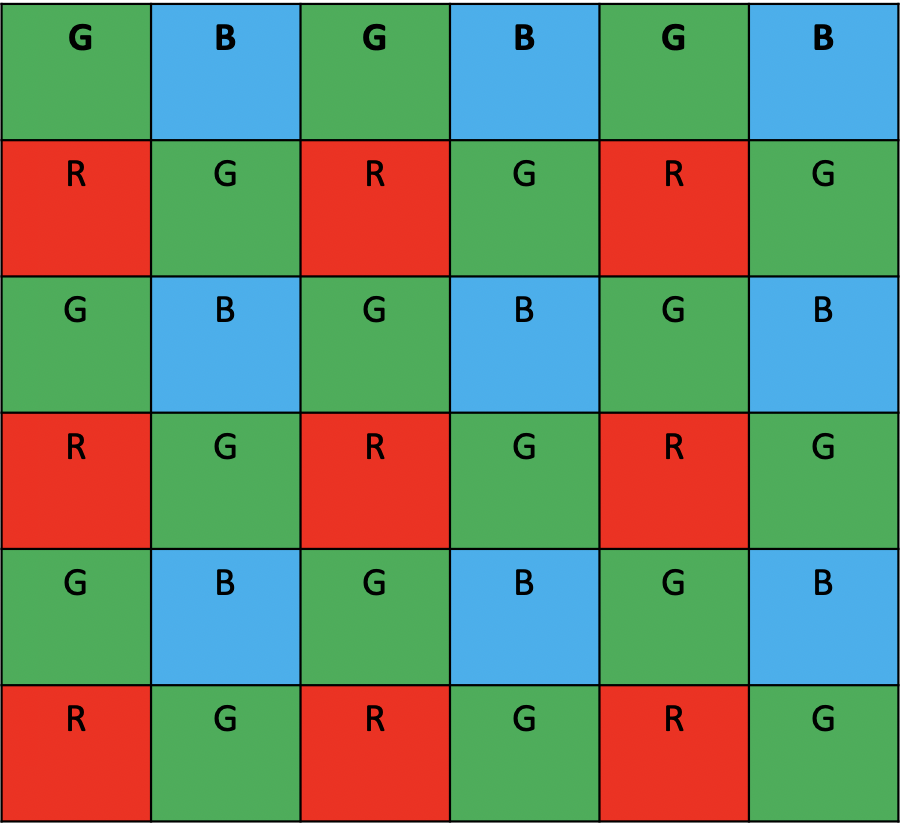}
  \caption{Example Bayer filter used for the image sensor in a digital camera. The label of each pixel is one of the three basic colors (red, green, or blue) to which it is sensitive. \cite{Bayer_filter}}
  \label{fig:Bayer_filter}
\end{figure}

The QDC generates the raw image as an array of size $2592 \times 1944 \times 1$; however, in order to represent an image in color space, each pixel needs information about 3 basic colors: red, green, and blue. In this simulation, color interpolation is performed in a way that mimics how color is created in real digital cameras. As described in \cite{Bayer_filter}, the array of photo sensors in modern digital cameras often contains three kinds of elements, each of which is sensitive to one basic color R, G, or B. These elements are arranged in a certain order called the Bayer Pattern, as shown in Figure \ref{fig:Bayer_filter}. As a result, each pixel has information about one color component that its neighbors are missing, and an interpolation, called demosaicing, is performed to obtain all three color components of each pixel.

We perform an independent linear interpolation on each color plane.
For a pixel defined by index $(i, j)$ with luminance $L_{i, j}$, if its label is green, then its color tuple, defined as $(R, G, B)$, is
\begin{align}
    (\frac{L_{i-1, j} + L_{i+1, j}}{2}, L_{i,j}, \frac{L_{i, j-1} + L_{i, j+1}}{2})
\end{align}
or, for another G label in the GBRG pattern (there are twice as many G pixels as each of R and B),
\begin{align}
    (\frac{L_{i, j-1} + L_{i, j+1}}{2}, L_{i,j}, \frac{L_{i-1, j} + L_{i+1, j}}{2})
\end{align}
If the label on the pixel is red, then its color tuple is
\begin{align}
    (L_{i,j}, \frac{L_{i-1, j} + L_{i+1, j} + L_{i, j-1} + L_{i, j+1}}{4}, \\ \frac{L_{i-1, j-1} + L_{i+1, j+1} + L_{i+1, j-1} + L_{i-1, j+1}}{4})
\end{align}
The color tuple for pixels with blue labels can be derived similarly using the equation for red pixels. As a result, each pixel now has information for all three color components, and a colored image array of size $2592 \times 1944 \times 3$ is obtained.

The next step in image processing is white balance. White balance is the process of removing unrealistic color casts due to the color temperature of the light source, so that white objects appear white in digital camera images. We perform white balancing by scaling the image to a color temperature of 5000 Kelvin. According to the black body color temperature table \cite{blackbody_color}, 5000 Kelvin correspond to the RGB color tuple (255, 213, 171).  Therefore, all pixels (R, G, B) in the image are multiplied by ($\frac{255}{255}$, $\frac{213}{255}$, $\frac{171}{255}$).


After white balance, background noise is added to the simulated images. Due to the proprietary nature of the electronics noise profile of the HTC Wildfire, instead of simulating electronic noise in section \ref{sec:allpix}, noise is added here using actual image data taken from an HTC Wildfire in dark conditions. Since the input to the CNN that is used to classify these images uses only a $64 \times 64$ pixel square centered on the brightest point, instead of choosing a whole data image as background, a $64 \times 64$ pixel region is randomly selected from a random data image and added to the cropped simulated image before being put into the CNN. Also, one thing worth noting here is that data taken by DECO on the HTC Wildfire used different ISO values due to changing software versions throughout the lifespan of the project. The ISO setting controls the sensitivity of the camera sensor, and in the simulation its effect has been integrated with the QDC conversion factor. Background noise levels in data events are sensitive to these changes. Since this simulation focuses on comparing simulated track events with track events in data, the backgrounds added to the simulated images need to be selected from data events with the same ISO values as the training data. Based on the EXIF header data of images collected by DECO, more than $95\%$ of the track events in data taken by HTC Wildfire have ISO value 881, so DECO images in data with this ISO value are collected as the background pool. During the image processing in simulation, a random background is drawn from the pool, a random $64 \times 64$ region is cropped from it, and this region is added to the cropped simulated image.

In the last step, a simulated image is put into the DECO CNN \cite{Winter:2018enx} for classification. The input to the CNN is a gray-scale image array of size $64 \times 64 \times 1$, so the simulated image needs to be further processed. As mentioned in the background noise addition section, the simulated image has been cropped to a $64 \times 64 \times 3$ image centered around its brightest pixel before adding background. Then the RGB colored image is converted into gray-scale by calculating the luminance as $L = 0.299R + 0.587G + 0.114B$ on each pixel. The resulting $64 \times 64 \times 1$ array is input to the CNN.  If its probability of being a track is larger than $50\%$, it will be classified by the CNN as a track event.

\subsection{Atmospheric Muon Parameterization}
\label{sec:MC_sim}

\begin{figure}[ht]
\begin{subfigure}{0.5\textwidth}
  \centering
  \includegraphics[scale = 0.23]{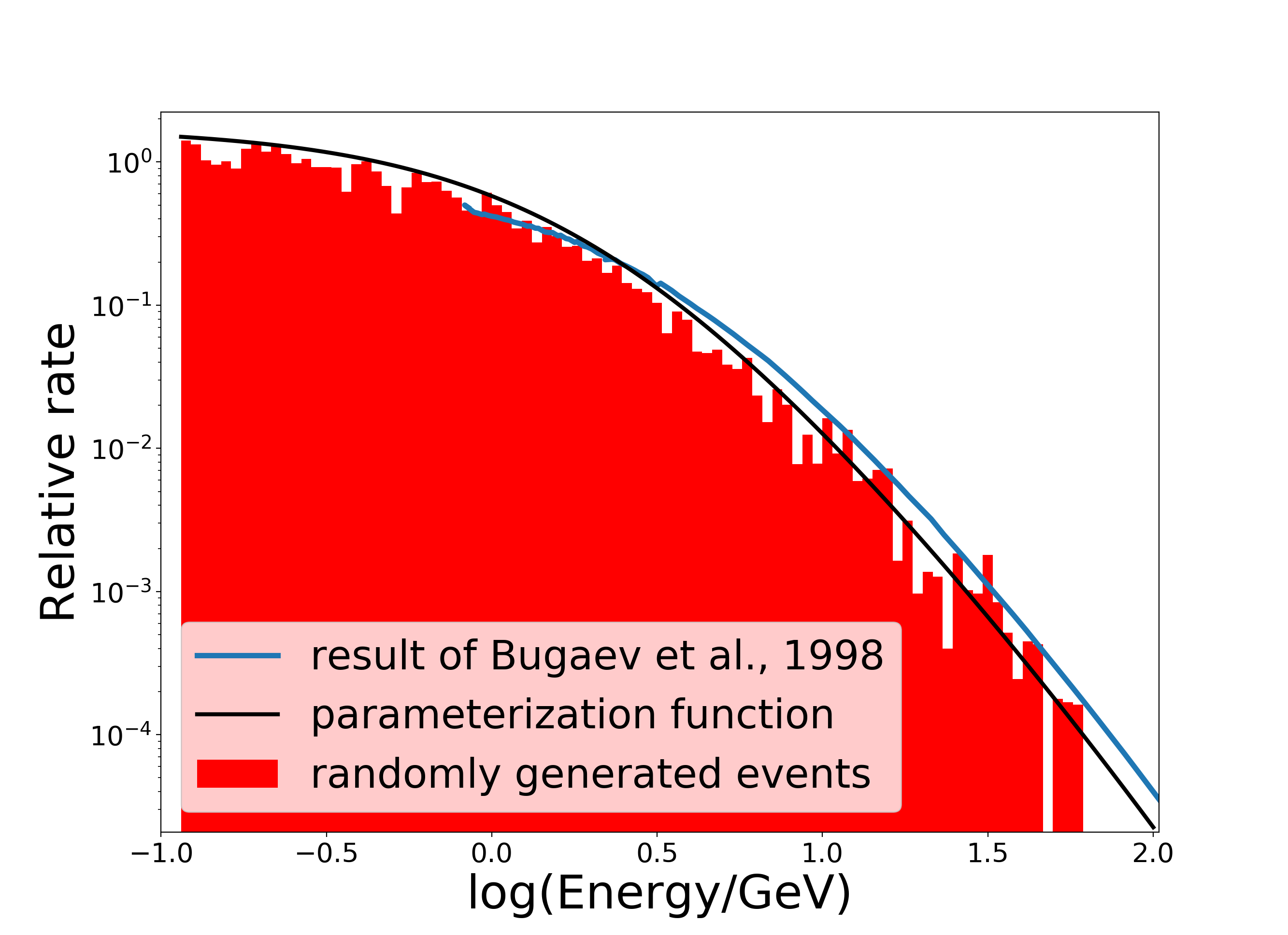}
  \caption{}
\end{subfigure}
\begin{subfigure}{0.5\textwidth}
  \centering
  \includegraphics[scale = 0.23]{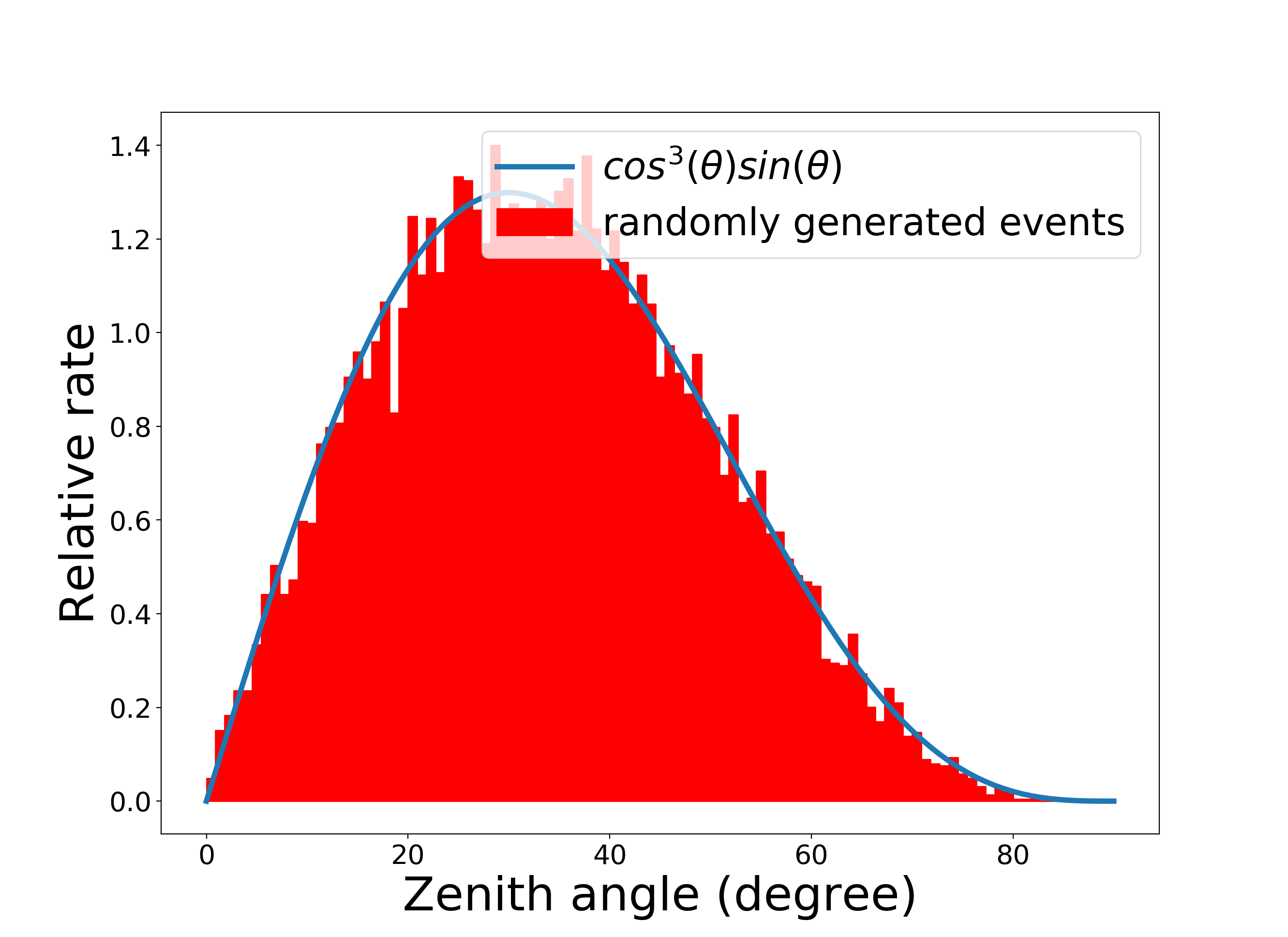}
  \caption{}
\end{subfigure}
\caption{Energy and zenith angle distributions of atmospheric muons. In figure (a) the vertical atmospheric muon energy distribution in \cite{muon_flux_pic} is compared against 3000 simulated events at $\theta = 0^{\circ}$ and the parameterization given as Equation~\ref{eq:muon_flux}. Note the distribution spans four orders of magnitude in rate.  In Figure (b) 3000 simulated events at E = 1 GeV are compared against the empirical atmospheric muon angular distribution formula $\cos^2\theta$, with an extra $\sin\theta \cos\theta$ accounting for the conversion from $d\Omega$ to $d\theta$ and the projection effect in detector effective area\cite{Vandenbroucke:2015kkk}.}
\label{fig:compare_muon_flux}
\end{figure}

Instead of just focusing on simulating individual events, this simulation performs a Monte Carlo simulation of cosmic-ray muons, since their flux can be well described by the atmospheric muon flux spectrum in Equation~\ref{eq:muon_flux} below (see e.g. \cite{muon_flux}) with an additional factor of $\cos\theta$ that represents the projection effect converting detector geometrical areay to effective area. 
A cross check between the expected atmospheric muon flux and 3000 randomly generated muon events is made. In Figure~\ref{fig:compare_muon_flux}(a), the vertical differential momentum spectra of conventional muons at sea level is extracted from \cite{muon_flux_pic}, and it is compared with both the parameterization function (Equation~\ref{eq:muon_flux}) and 3000 muons randomly generated by this MC simulation with $\theta=0$. In Figure~\ref{fig:compare_muon_flux}(b), the empirical approximation of the low energy atmospheric muon zenith angle distribution is used to compare with the angular distribution of 3000 muons randomly generated at fixed energy 1 GeV.

\begin{equation}
    \begin{multlined}
    \frac{dN}{dEd\Omega dAdt} = E^{-2.7} (\frac{1}{1 + 1.11 \frac{E cos\theta}{115 GeV}} + \frac{0.054}{1 + 1.11\frac{E cos\theta}{850 GeV}}) \times \\
    (\frac{E}{E + \frac{2 GeV}{cos \theta}})^{2.7}  cos\theta
    \end{multlined}
    \label{eq:muon_flux}
\end{equation}

In this Monte Carlo simulation, the position where test particles are generated is fixed to be the center of the CMOS top surface. Under the assumption that all pixels in the CMOS share the same image processing procedure, the exact position where a test particle hits the CMOS does not matter. This assumption is not always true, since as discussed in~\cite{gain_factor}, a technique called lens-shading, which assigns pixels different gain factors to compensate for the decreased intensity of light near the edges of a flat sensor, is common in digital cameras. For the sample Galaxy S7 phone used in~\cite{gain_factor}, the gain factor varies by roughly a factor of three. In our simulation of an HTC Wildfire in DECO, the lens shading gain factor has been ignored, but it might be of interest for future study.

\section{Simulation Results}
\label{sec:results}

\subsection{Deposited Energy}
\label{sec:energy_rec}

During the simulation of charged particles in the CMOS described in section \ref{sec:allpix}, we have the charge on pixel map before it is put into the QDC and any further image processing procedures are applied. Using the summed charge on pixels multiplied by the charge creation energy 3.62 eV listed in table \ref{tab:sim_parameter}, the energy loss of charged particles inside the CMOS can be calculated. Although this exact procedure works only for simulated images and not data events, since it uses the deposited charge before processing, it serves as a cross check of our simulation if we compare the deposited energy against the ideal energy loss calculated by integrating the Bethe-Bloch formula. Although we work directly with charge for this simulation validation, the conversion from charge to RGB image, described above as applied to simulation data, could be inverted and applied to experimental data in order to measure physical deposited energy in experimental DECO data.  Furthermore, retaining RAW format image data could retain information closer to the original depsoited charge in experimental data.

For muons and electrons (which always interact in the sensor), at each of the different initial kinetic energy levels from 10 keV to 100 GeV, 500 test particles are simulated with initial zenith angle $45^\circ$. For photons (which can have a small probability of interacting within the sensor, depending on their energy), at each of the different initial kinetic energy levels from 10 keV to 100 GeV, a large enough number of test particles are simulated with initial zenith angle $45^\circ$ to make sure 500 events are collected at each energy level. Then the charge deposited is summed over pixels and converted into the original energy lost to ionization by multiplying by 3.62 eV.

The ideal energy loss from integrating the Bethe-Bloch formula is calculated in the following manner.  First, we assume that the particle trajectory is straight without any deflection.  Then, based on the initial zenith angle and CMOS depletion thickness, the length of this ideal trajectory can be calculated.  Next, with the known initial energy of the test particle, we can numerically integrate the Bethe-Bloch formula along the path to get the total energy deposited. If during such integration the particle kinetic energy reaches zero, then the total energy lost is set to its initial kinetic energy.

\begin{figure}[h]
\centering
  \includegraphics[width=0.55\textwidth]{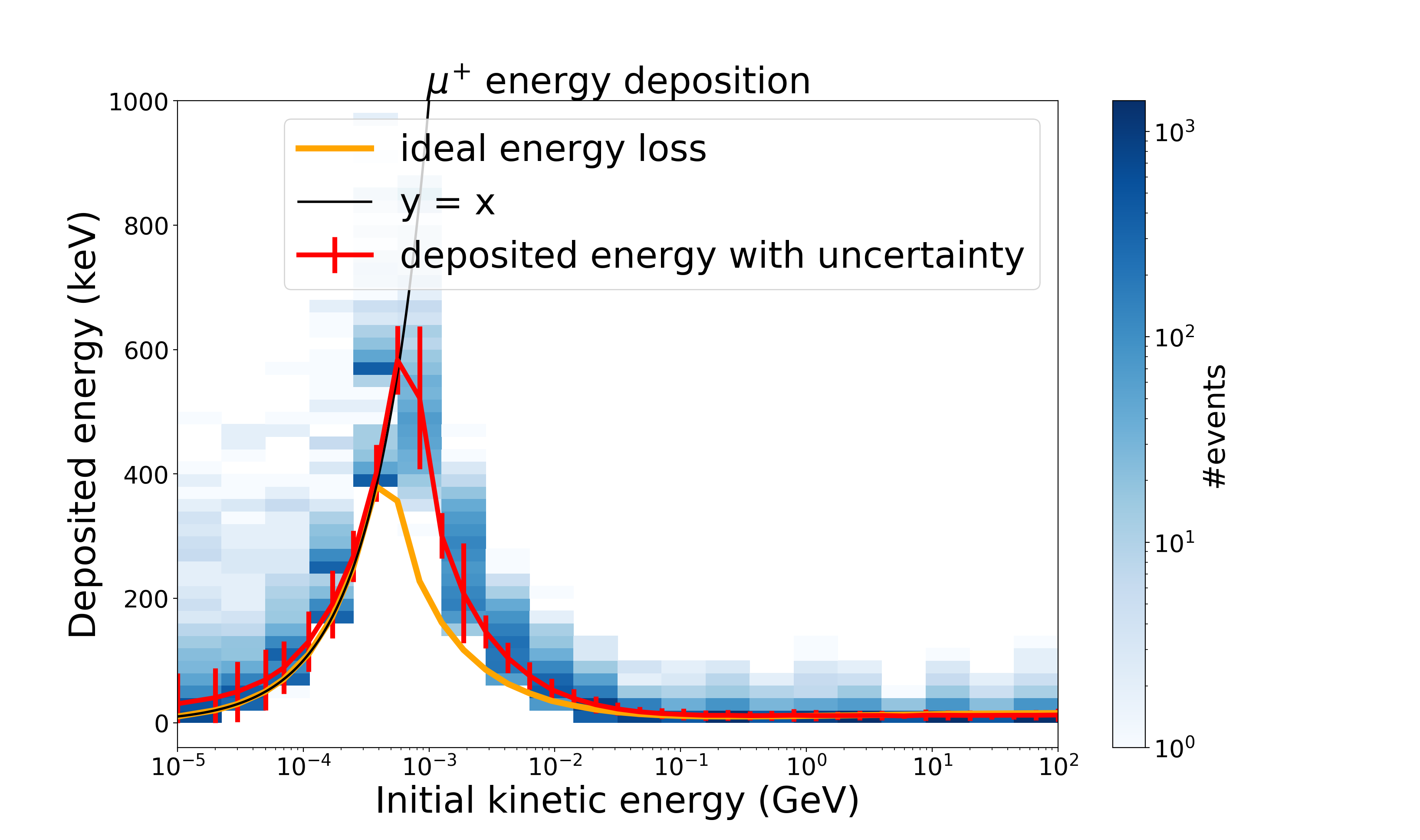}
  \caption{Total muon deposited energy in simulation compared with ideal energy loss from integration of the Bethe-Bloch formula. The mean and standard deviation of deposited energy is computed from the histogram of all simulated events, which is shown in blue.}
  \label{fig:muon_energy_cali}
\end{figure}

For muons, the result of comparing deposited energy against ideal energy loss as a function of initial kinetic energy is shown in Figure~\ref{fig:muon_energy_cali}. The orange line represents the ideal energy loss according to Bethe-Bloch, which in the beginning is simply $y = x$ since muons at low energy are often completely absorbed within the CMOS. The blue histogram represents the distribution of deposited energy for all test particles at different initial kinetic energies. The red line is the mean and standard deviation calculated from the blue histogram.

At low energy, the deposited energy distribution exhibits a tail extending to greater energy loss than that predicted by ionization loss alone. This higher energy loss is due to muon decay. As the initial muon energy decreases, the muon becomes more likely to stop and decay within the sensor, resulting in a high-energy electron. The energy deposited by these ``Michel electrons'' makes the total energy loss higher than expected from muon ionization alone. At medium energy levels, around 1 MeV, the ionization loss of muons in silicon is greatest.  Between $\sim0.3$~MeV and $\sim$10 MeV, we see the largest disagreement between deposited energy and the analytical estimation of ionization energy loss. This is likely due to multiple Coulomb scattering: in tihs energy range, the muon is too energetic to be absorbed within the sensor, but is not energetic enough to follow a straight trajectory through it.  Therefore it deposits more energy than expected from the analytical approximation, which assumes ionization loss along a straight line through the sensor.  There is also the largest standard deviation of the deposited energy in this range. Above $\sim$10 MeV, muons are minimum ionizing particles which have a more stable energy loss rate and are less likely to experience scattering. As a result, the agreement between deposited energy and idealized energy loss is good.

\begin{figure}[h]
\centering
  \includegraphics[width=0.55\textwidth]{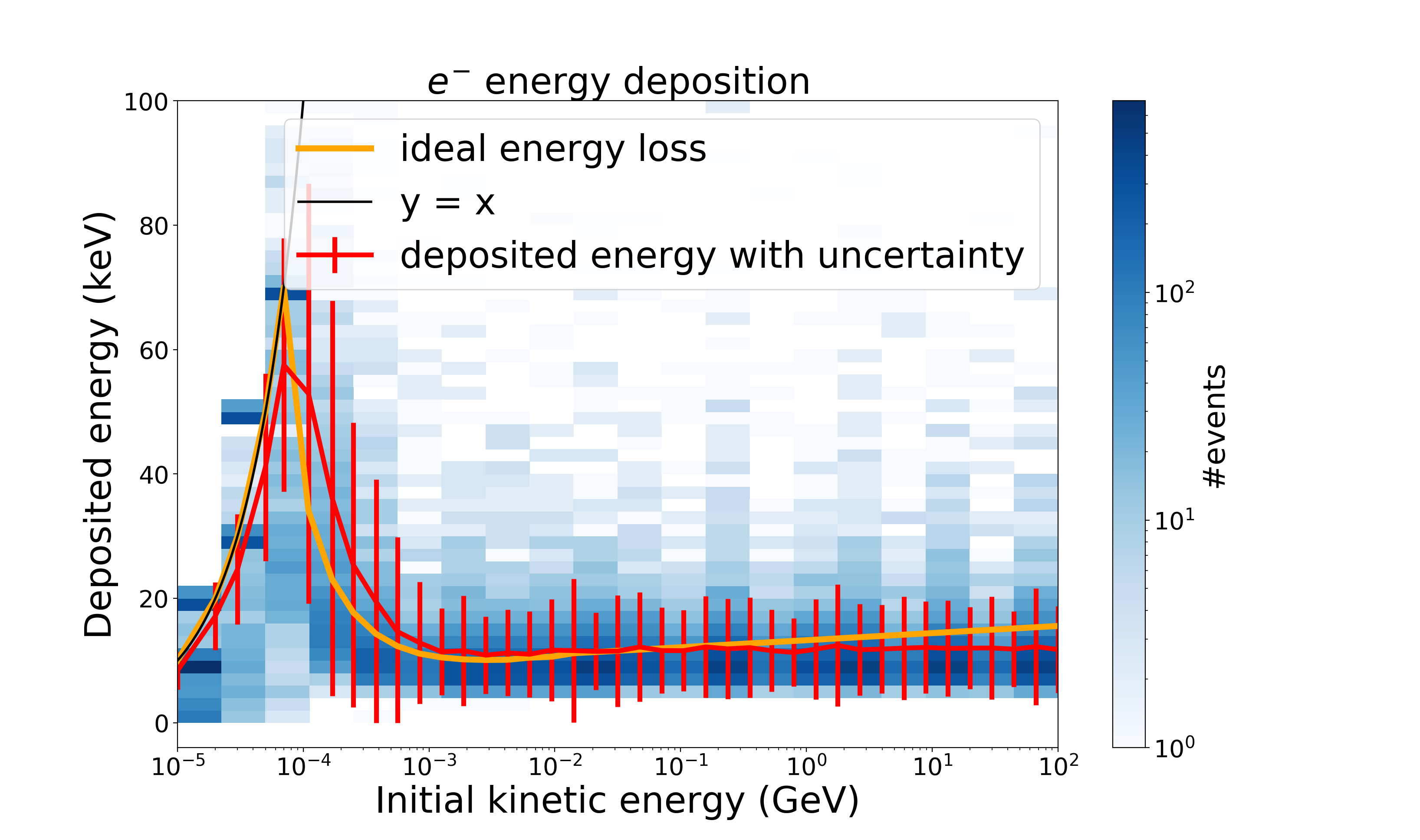}
  \caption{Total electron deposited energy in simulation compared with ideal energy loss from integration. The mean and standard deviation of deposited energy is computed from the histogram of all simulated events.}
  \label{fig:electron_energy_cali}
\end{figure}

For electrons, a similar comparison between simulation results and analytical expectations is shown in Figure \ref{fig:electron_energy_cali}. At energies smaller than $\sim$1 MeV, the difference between deposited energy and ideal energy loss is caused by multiple Coulomb scattering that shifts electrons away from their initial trajectories, similar to that seen for muons. For initial kinetic energies greater than $\sim$1 MeV, electrons become minimum ionizing particles and are less likely to be deflected. As a result, the agreement between deposited energy and ideal energy loss is very good. 

\begin{figure}[h!]
\begin{subfigure}{0.5\textwidth}
  \centering
  \includegraphics[scale = 0.23]{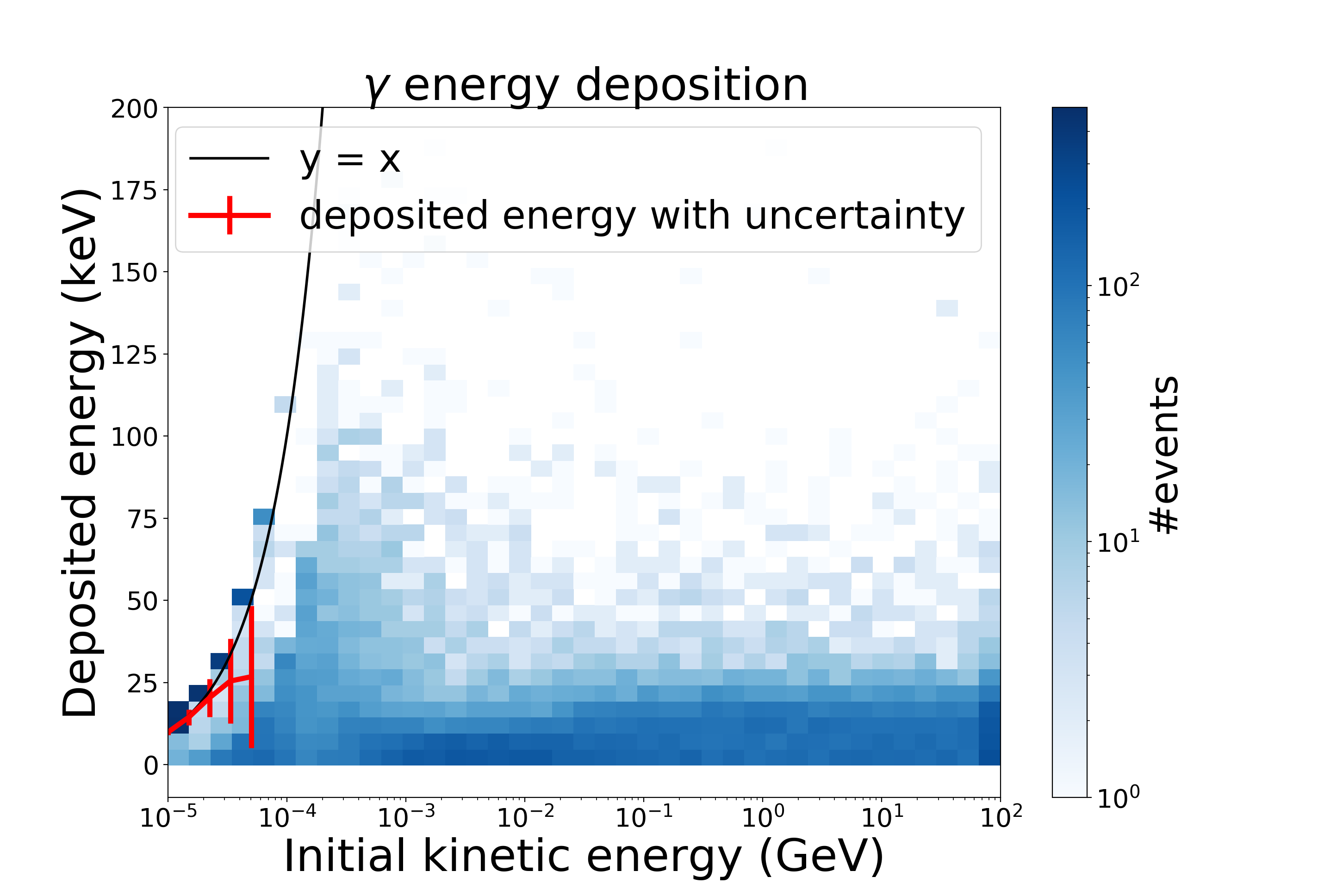}
  \caption{}
\end{subfigure}
\begin{subfigure}{0.5\textwidth}
  \centering
  \includegraphics[scale = 0.23]{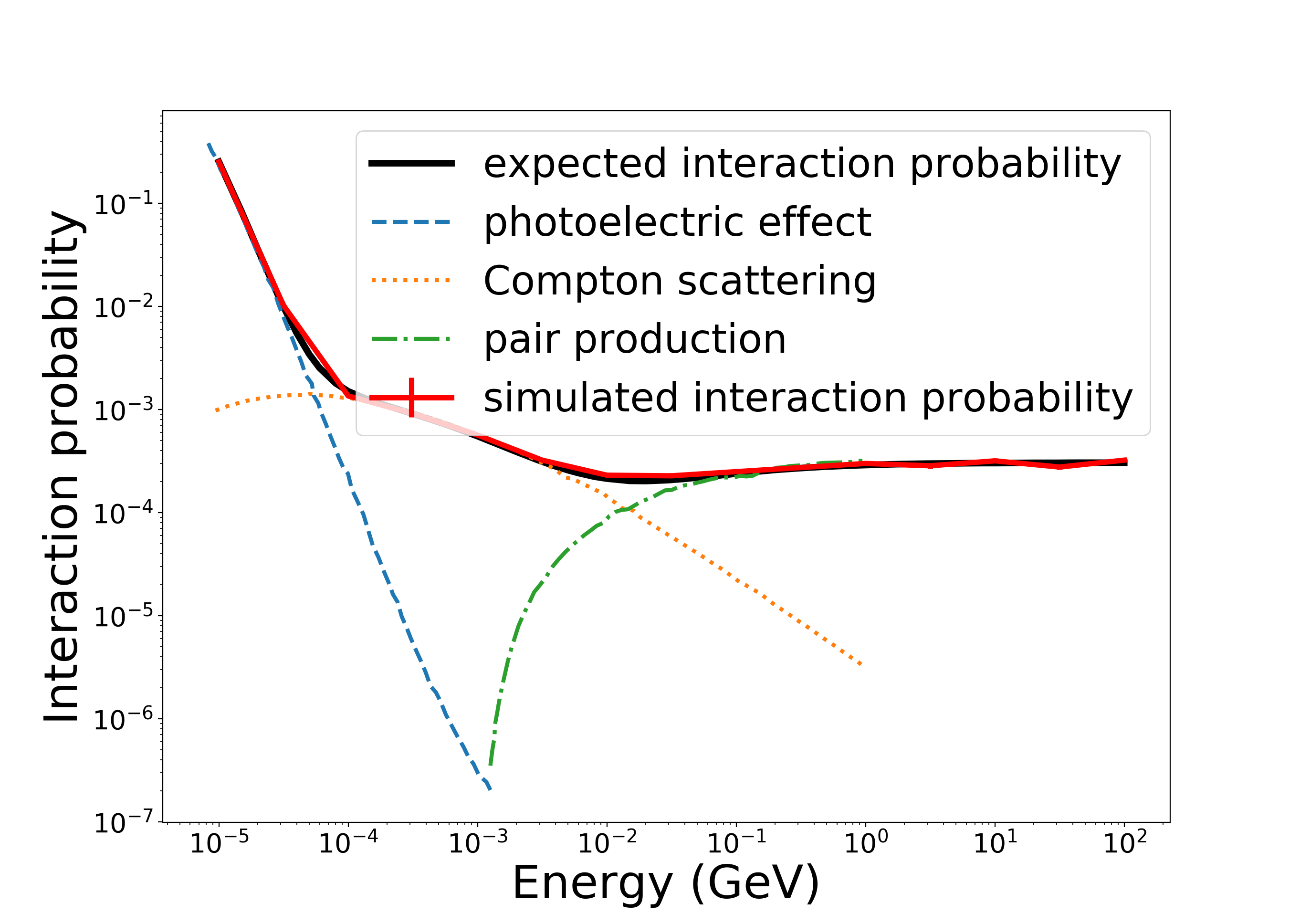}
  \caption{}
\end{subfigure}
\caption{Photon deposited energy and interaction probability. In Figure (a) the total deposited energy in simulation is plotted. The mean and standard deviation of deposited energy is calculated for the photoelectric effect regime. Figure (b) shows the photon interaction probability derived from the simulation results compared to the analytical expectation based on photon interaction cross sections in \cite{photon_crosssec}.}
\label{fig:gamma_energy_cali}
\end{figure}

Results for photons are shown in Figure~\ref{fig:gamma_energy_cali}. As discussed above, unlike electrons and muons, photons are unlikely to interact within the silicon sensor.  At each of the different initial photon energy levels from 10 keV to 100 GeV, a large enough number of test particles are simulated with initial zenith angle $45^\circ$ to ensure that at least 500 events interact in the sensor.  Below $\sim$100~keV, the photon interaction probability (dominated by the photo-electric effect) is greatest, and the resulting electron is likely to be absorbed entirely within the sensor.  This is confirmed by the large fraction of events with deposited energy equal to initial photon energy, following the $y=x$ line in Figure~\ref{fig:gamma_energy_cali}.  At $\sim$100~keV, the outgoing electron is more likely to escape the sensor than at lower energies, resulting in mean deposited energy less than the initial photon energy.  Above $\sim$100~keV, many events do not interact at all, but a fraction undergo Compton scattering or pair production.  The probability of interaction derived from the Monte Carlo results is plotted in Figure~\ref{fig:gamma_energy_cali} (b), and it agrees well with the analytical interaction probability calculated from the photon interaction cross sections in \cite{photon_crosssec} in all three regimes (hoto-electric, Compton, and pair).

\subsection{Comparison of Individual Images}
\label{sec:single_comp}

\begin{figure*}[h!]

\begin{subfigure}{0.3333\textwidth}
  \centering
  \includegraphics[scale = 0.3]{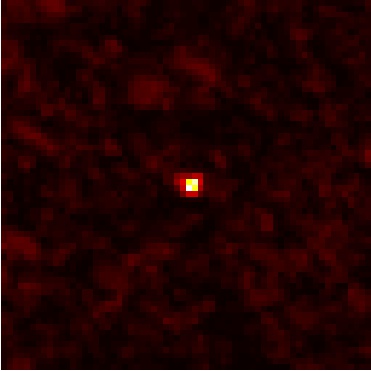}
  \captionsetup{justification=centering}
  \caption{T = 0, W = 0,\\ S = 1}
\end{subfigure}
\begin{subfigure}{0.3333\textwidth}
  \centering
  \includegraphics[scale = 0.3]{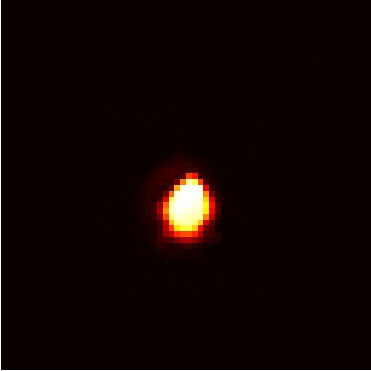}
  \captionsetup{justification=centering}
  \caption{T = 0, W = 0,\\ S = 1}
\end{subfigure}
\begin{subfigure}{0.3333\textwidth}
  \centering
  \includegraphics[scale = 0.3]{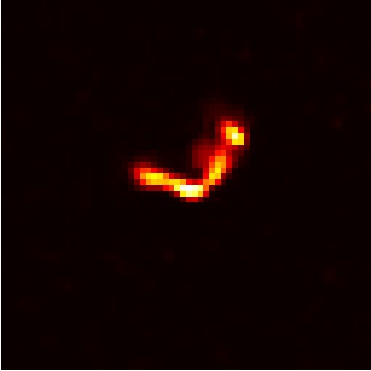}
  \captionsetup{justification=centering}
  \caption{T = 0, W = 1,\\ S = 0}
\end{subfigure}

\begin{subfigure}{0.3333\textwidth}
  \centering
  \includegraphics[scale= 0.3]{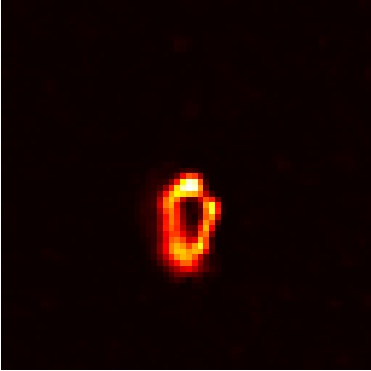}
  \captionsetup{justification=centering}
  \caption{T = 0, W = 1,\\ S = 0}
\end{subfigure}
\begin{subfigure}{0.3333\textwidth}
  \centering
  \includegraphics[scale = 0.3]{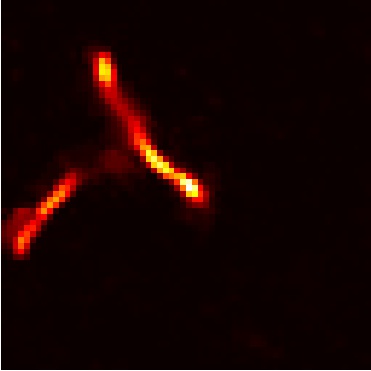}
  \captionsetup{justification=centering}
  \caption{T = 0.267, W = 0.733,\\ S = 0}
\end{subfigure}
\begin{subfigure}{0.3333\textwidth}
  \centering
  \includegraphics[scale = 0.3]{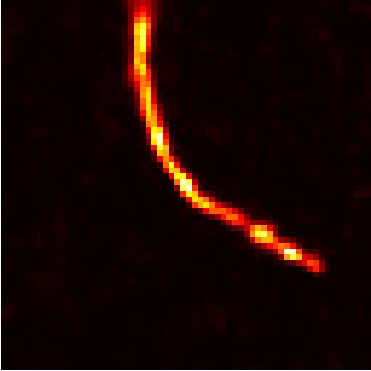}
  \captionsetup{justification=centering}
  \caption{T = 0.384, W = 0.613,\\ S = 0.002}
\end{subfigure}

\begin{subfigure}{0.3333\textwidth}
  \centering
  \includegraphics[scale = 0.3]{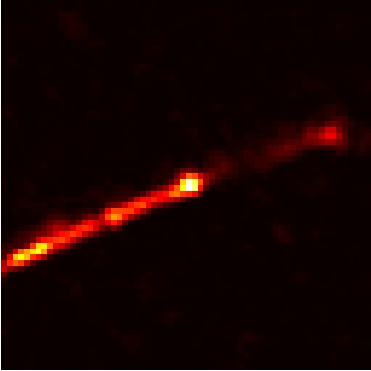}
  \captionsetup{justification=centering}
  \caption{T = 0.626, W = 0.368,\\ S = 0.003}
\end{subfigure}
\begin{subfigure}{0.3333\textwidth}
  \centering
  \includegraphics[scale = 0.3]{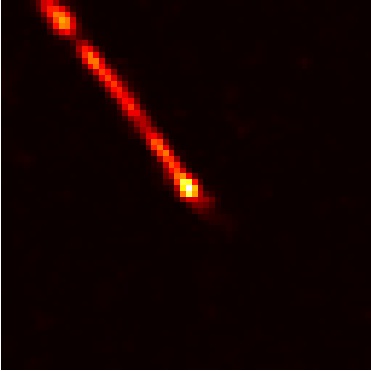}
  \captionsetup{justification=centering}
  \caption{T = 0.951, W = 0.048,\\ S = 0}
\end{subfigure}
\begin{subfigure}{0.3333\textwidth}
  \centering
  \includegraphics[scale = 0.3]{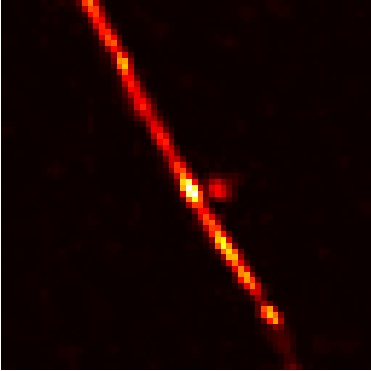}
  \captionsetup{justification=centering}
  \caption{T = 0.967, W = 0.033,\\ S = 0}
\end{subfigure}

\begin{subfigure}{0.3333\textwidth}
  \centering
  \includegraphics[scale = 0.3]{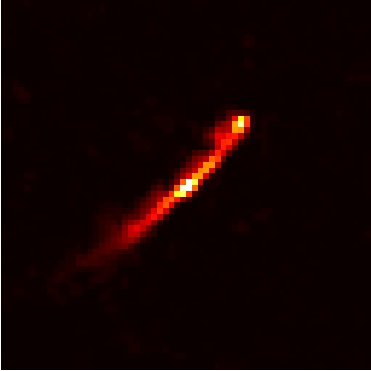}
  \captionsetup{justification=centering}
  \caption{T = 0.970, W = 0.030,\\ S = 0}
\end{subfigure}
\begin{subfigure}{0.3333\textwidth}
  \centering
  \includegraphics[scale = 0.3]{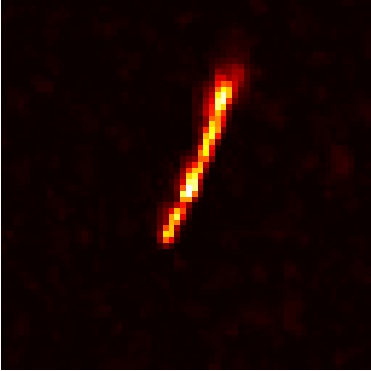}
  \captionsetup{justification=centering}
  \caption{T = 0.974, W = 0.026,\\ S = 0}
\end{subfigure}
\begin{subfigure}{0.3333\textwidth}
  \centering
  \includegraphics[scale = 0.3]{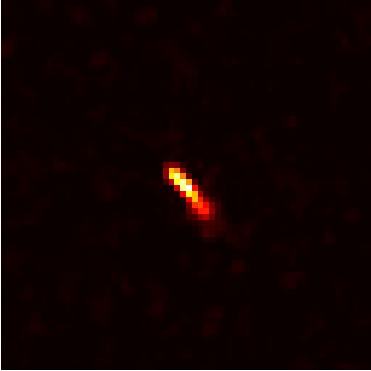}
  \captionsetup{justification=centering}
  \caption{T = 0.985, W = 0.015,\\ S = 0}
\end{subfigure}

\caption{Example experimental data events collected by DECO. The CNN classified probability of each image is labeled as T (track), W (worm), and S (spot).}
\label{fig:real_image}
\end{figure*}

In this section, example images from both experimental events observed by DECO and simulated events with different particle types and bias voltages are shown with their probability of being classified as track, worm, and spot by the CNN. Example experimental DECO data events are shown in Figure \ref{fig:real_image}. As discussed in Section~\ref{sec:intro}, the first two events are classified as spots, since they have small, approximately circular clusters of pixels. Figure~\ref{fig:real_image} (c) to (f) have a large probability of being worms, since they show curved clusters of pixels. Figure \ref{fig:real_image} (g) to (l) have a long straight cluster of pixels, so they are classified with large probability of being tracks.

\begin{figure*}[h!]
\begin{subfigure}{0.3333\textwidth}
  \centering
  \includegraphics[scale = 0.3]{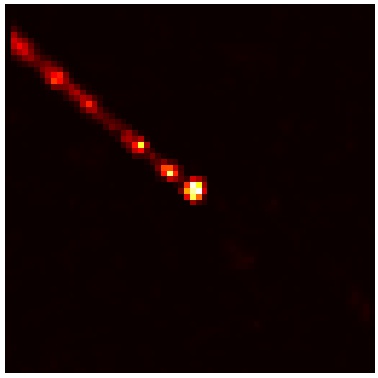}
  \captionsetup{justification=centering}
  \caption{100 keV, $\theta$ = $20^\circ$, T = 0.741,\\ W = 0.257, S = 0.002}
\end{subfigure}
\begin{subfigure}{0.3333\textwidth}
  \centering
  \includegraphics[scale = 0.3]{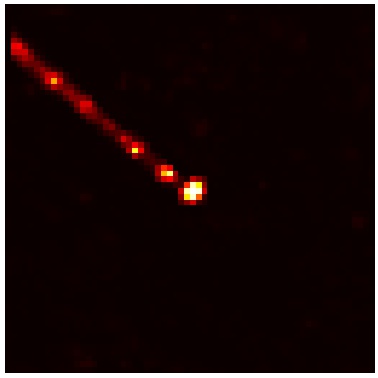}
  \captionsetup{justification=centering}
  \caption{100 keV, $\theta$ = $50^\circ$, T = 0.655,\\ W = 0.342, S = 0.003}
\end{subfigure}
\begin{subfigure}{0.3333\textwidth}
  \centering
  \includegraphics[scale = 0.3]{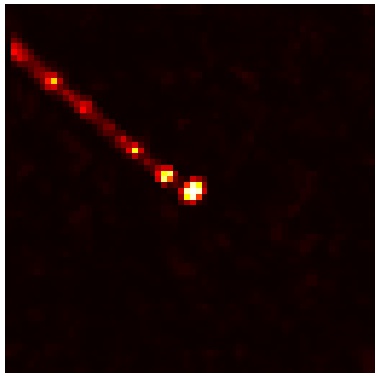}
  \captionsetup{justification=centering}
  \caption{100 keV, $\theta$ = $80^\circ$, T = 0.506,\\ W = 0.492, S = 0.003}
\end{subfigure}

\begin{subfigure}{0.3333\textwidth}
  \centering
  \includegraphics[scale = 0.3]{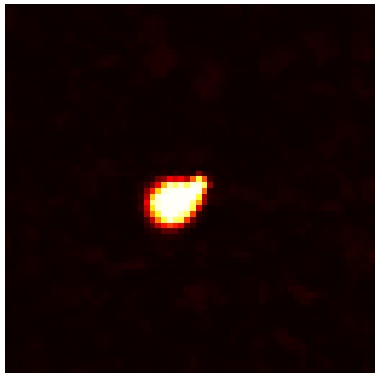}
  \captionsetup{justification=centering}
  \caption{1 MeV, $\theta$ = $20^\circ$, T = 0.005,\\ W = 0.009, S = 0.976}
\end{subfigure}
\begin{subfigure}{0.3333\textwidth}
  \centering
  \includegraphics[scale = 0.3]{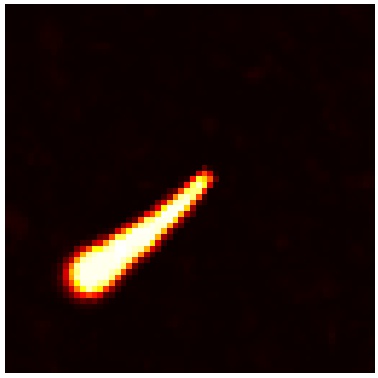}
  \captionsetup{justification=centering}
  \caption{1 MeV, $\theta$ = $50^\circ$, T = 0.701,\\ W = 0.295, S = 0}
\end{subfigure}
\begin{subfigure}{0.3333\textwidth}
  \centering
  \includegraphics[scale = 0.3]{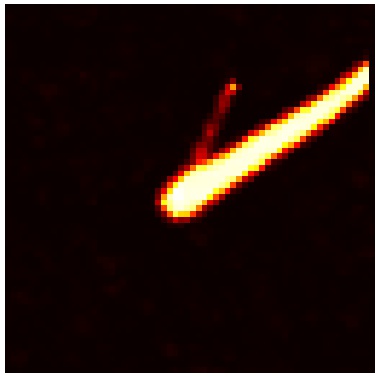}
  \captionsetup{justification=centering}
  \caption{1 MeV, $\theta$ = $80^\circ$, T = 0.175,\\ W = 0.645, S = 0}
\end{subfigure}

\begin{subfigure}{0.3333\textwidth}
  \centering
  \includegraphics[scale = 0.3]{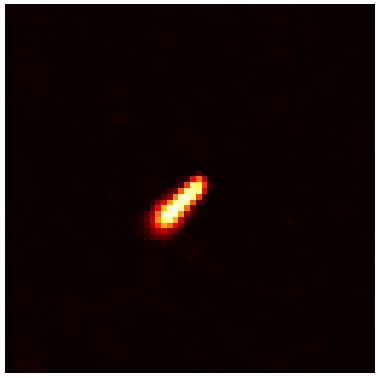}
  \captionsetup{justification=centering}
  \caption{10 MeV, $\theta$ = $20^\circ$, T = 0.837,\\ W = 0.161, S = 0.002}
\end{subfigure}
\begin{subfigure}{0.3333\textwidth}
  \centering
  \includegraphics[scale = 0.3]{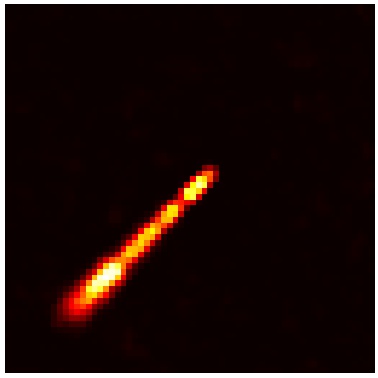}
  \captionsetup{justification=centering}
  \caption{10 MeV, $\theta$ = $50^\circ$, T = 0.998,\\ W = 0.018, S = 0}
\end{subfigure}
\begin{subfigure}{0.3333\textwidth}
  \centering
  \includegraphics[scale = 0.3]{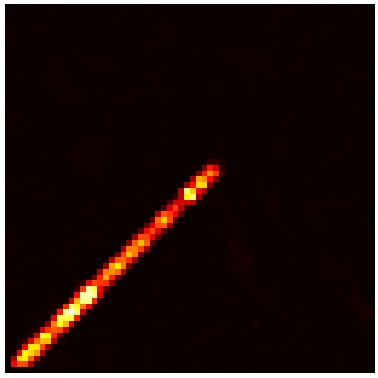}
  \captionsetup{justification=centering}
  \caption{10 MeV, $\theta$ = $80^\circ$, T = 0.998,\\ W = 0.002, S = 0}
\end{subfigure}

\begin{subfigure}{0.3333\textwidth}
  \centering
  \includegraphics[scale = 0.3]{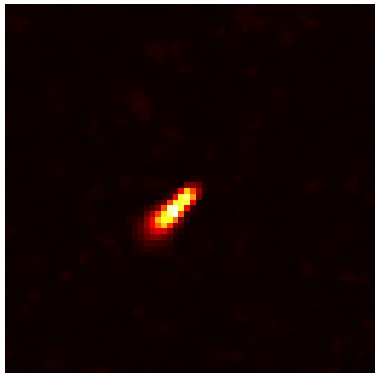}
  \captionsetup{justification=centering}
  \caption{100 MeV, $\theta$ = $20^\circ$, T = 0.908,\\ W = 0.089, S = 0.002}
\end{subfigure}
\begin{subfigure}{0.3333\textwidth}
  \centering
  \includegraphics[scale = 0.3]{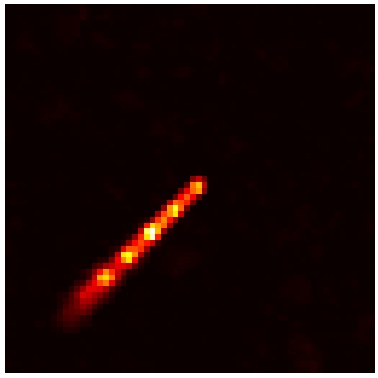}
  \captionsetup{justification=centering}
  \caption{100 MeV, $\theta$ = $50^\circ$, T = 1,\\ W = 0, S = 0}
\end{subfigure}
\begin{subfigure}{0.3333\textwidth}
  \centering
  \includegraphics[scale = 0.3]{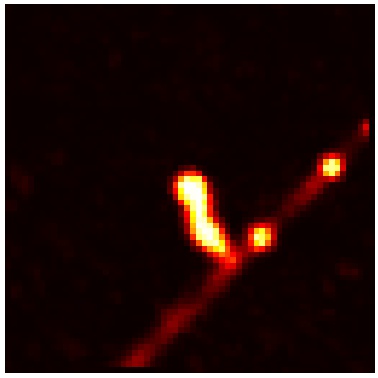}
  \captionsetup{justification=centering}
  \caption{100 MeV, $\theta$ = $80^\circ$, T = 0.017,\\ W = 0.983, S = 0}
\end{subfigure}

\begin{subfigure}{0.3333\textwidth}
  \centering
  \includegraphics[scale = 0.3]{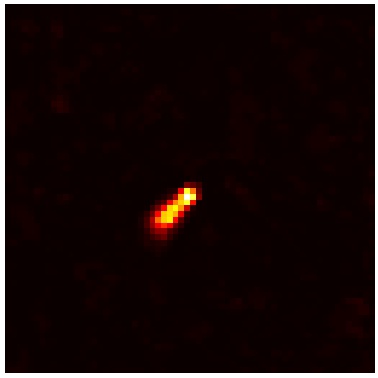}
  \captionsetup{justification=centering}
  \caption{1 GeV, $\theta$ = $20^\circ$, T = 0.845,\\ W = 0.13, S = 0.024}
\end{subfigure}
\begin{subfigure}{0.3333\textwidth}
  \centering
  \includegraphics[scale = 0.3]{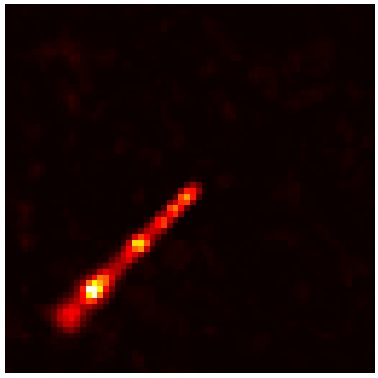}
  \captionsetup{justification=centering}
  \caption{1 GeV, $\theta$ = $50^\circ$, T = 0.995,\\ W = 0.005, S = 0}
\end{subfigure}
\begin{subfigure}{0.3333\textwidth}
  \centering
  \includegraphics[scale = 0.3]{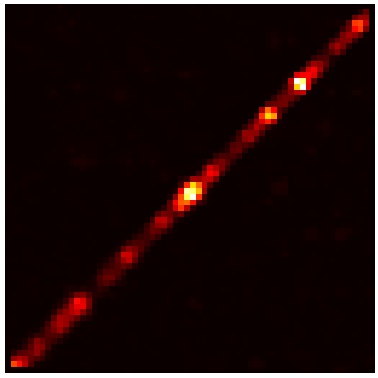}
  \captionsetup{justification=centering}
  \caption{1 GeV, $\theta$ = $80^\circ$, T = 0.999,\\ W = 0.001, S = 0}
\end{subfigure}

\caption{Example simulated muon events. The CNN classified probability is labeled as T (track), W (worm), and S (spot). Bias voltage is set to be -25 V, and $\phi$ is set to be $45^\circ$ in this simulation.}
\label{fig:sim_mu_25V}
\end{figure*}

\begin{figure*}[h!]
\begin{subfigure}{0.3333\textwidth}
  \centering
  \includegraphics[scale = 0.3]{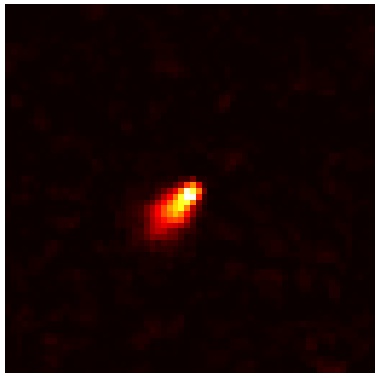}
  \captionsetup{justification=centering}
  \caption{100 MeV, $\theta$ = $20^\circ$, T = 0.153,\\ W = 0.393, S = 0.447}
\end{subfigure}
\begin{subfigure}{0.3333\textwidth}
  \centering
  \includegraphics[scale = 0.3]{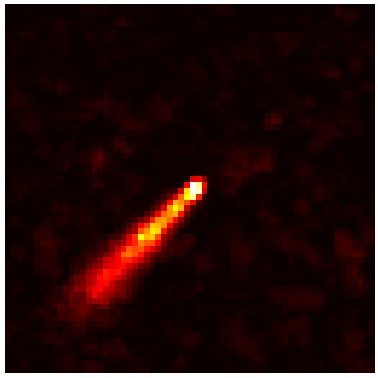}
  \captionsetup{justification=centering}
  \caption{100 MeV, $\theta$ = $50^\circ$, T = 0.816,\\ W = 0.184, S = 0}
\end{subfigure}
\begin{subfigure}{0.3333\textwidth}
  \centering
  \includegraphics[scale = 0.3]{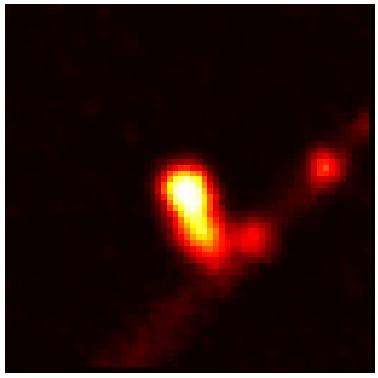}
  \captionsetup{justification=centering}
  \caption{100 MeV, $\theta$=$80^\circ$, T = 0,\\ W = 0, S = 0}
\end{subfigure}

\begin{subfigure}{0.3333\textwidth}
  \centering
  \includegraphics[scale = 0.3]{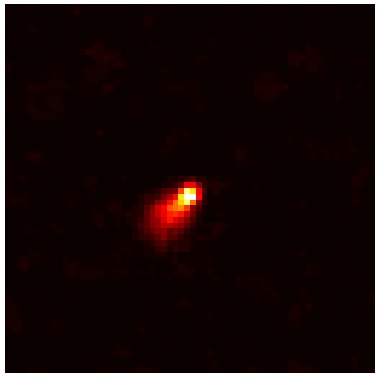}
  \captionsetup{justification=centering}
  \caption{1 GeV, $\theta$ = $20^\circ$, T = 0.056,\\ W = 0.085, S = 0.853}
\end{subfigure}
\begin{subfigure}{0.3333\textwidth}
  \centering
  \includegraphics[scale = 0.3]{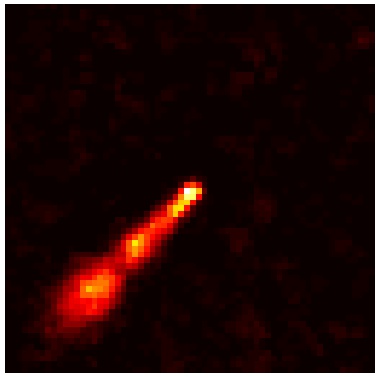}
  \captionsetup{justification=centering}
  \caption{1 GeV, $\theta$ = $50^\circ$, T = 0.699,\\ W = 0.295, S = 0.003}
\end{subfigure}
\begin{subfigure}{0.3333\textwidth}
  \centering
  \includegraphics[scale = 0.3]{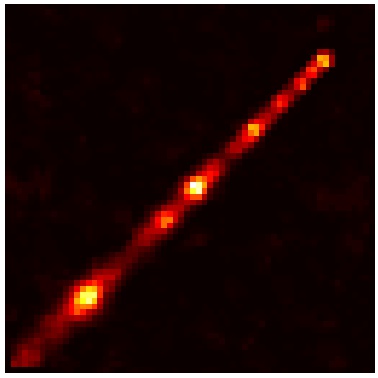}
  \captionsetup{justification=centering}
  \caption{1 GeV, $\theta$ = $80^\circ$, T = 0.982,\\ W = 0.018, S = 0}
\end{subfigure}

\caption{Example of simulated muon events. The CNN classified probability is labeled as T (track), W (worm), and S (spot). Bias voltage is set to be -5 V, and $\phi$ is set to be $45^\circ$ in this simulation.}
\label{fig:sim_mu_5V}
\end{figure*}

\begin{figure*}[h!]
\begin{subfigure}{0.3333\textwidth}
  \centering
  \includegraphics[scale = 0.3]{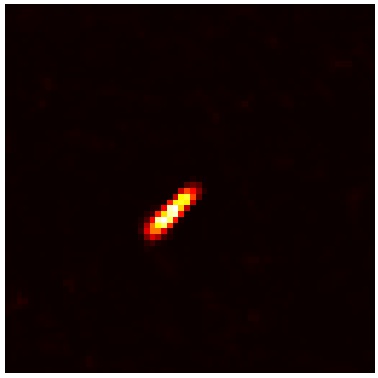}
  \captionsetup{justification=centering}
  \caption{100 MeV, $\theta$ = $20^\circ$, T = 0.981,\\ W = 0.019, S = 0}
\end{subfigure}
\begin{subfigure}{0.3333\textwidth}
  \centering
  \includegraphics[scale = 0.3]{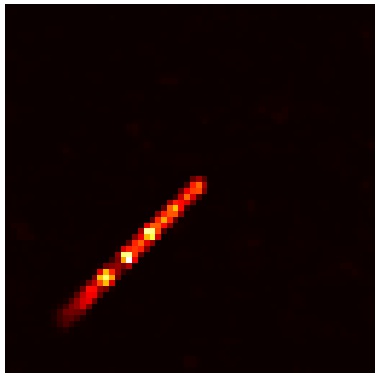}
  \captionsetup{justification=centering}
  \caption{100 MeV, $\theta$ = $50^\circ$, T = 1,\\ W = 0, S = 0}
\end{subfigure}
\begin{subfigure}{0.3333\textwidth}
  \centering
  \includegraphics[scale = 0.3]{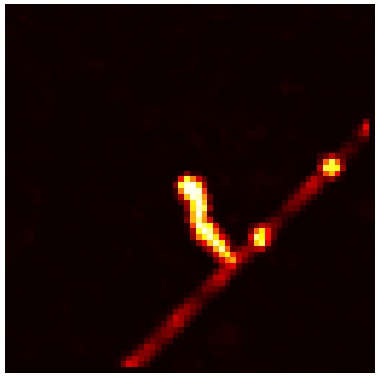}
  \captionsetup{justification=centering}
  \caption{100 MeV, $\theta$ = $80^\circ$, T = 0.06,\\ W = 0.940, S = 0}
\end{subfigure}

\begin{subfigure}{0.3333\textwidth}
  \centering
  \includegraphics[scale = 0.3]{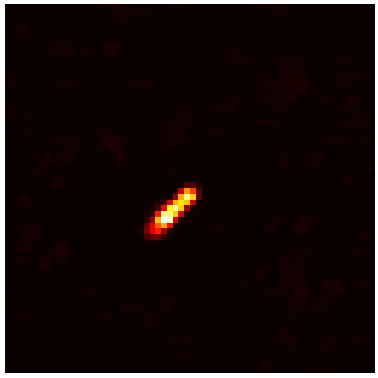}
  \captionsetup{justification=centering}
  \caption{1 GeV, $\theta$ = $20^\circ$, T = 0.970,\\ W = 0.030, S = 0}
\end{subfigure}
\begin{subfigure}{0.3333\textwidth}
  \centering
  \includegraphics[scale = 0.3]{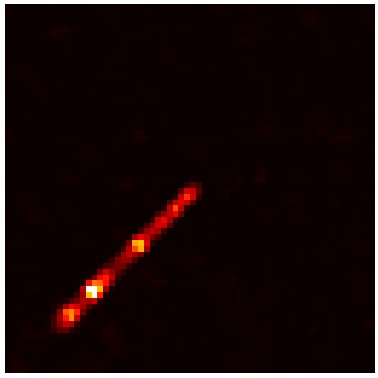}
  \captionsetup{justification=centering}
  \caption{1 GeV, $\theta$ = $50^\circ$, T = 0.998,\\ W = 0.002, S = 0}
\end{subfigure}
\begin{subfigure}{0.3333\textwidth}
  \centering
  \includegraphics[scale = 0.3]{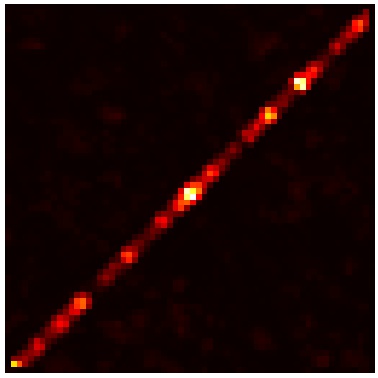}
  \captionsetup{justification=centering}
  \caption{1 GeV, $\theta$ = $80^\circ$, T = 1,\\ W = 0, S = 0}
\end{subfigure}

\caption{Example simulated muon events. The CNN classified probability is labeled as T (track), W (worm), and S (spot). Bias voltage is set to be -100 V, and $\phi$ is set to be $45^\circ$ in this simulation.}
\label{fig:sim_mu_100V}
\end{figure*}

Example simulated muon events, including CNN classification results, are shown in Figure~\ref{fig:sim_mu_25V}. In this image collection, each row has the same energy but different zenith angle including $20^\circ$, $50^\circ$, and $80^\circ$. In each row with the same initial energy, the same random seed is used for all three zenith angles to remove the event-to-event variation in stochastic interactions and focus only on the zenith angle effect. Also, each column keeps the same zenith angle but different energy levels selected uniformly in log(energy) from 100 keV to 1 GeV.  This range spans the full variety of behavior from muon decay to minimum ionization. In all simulations, the azimuth angle is fixed at $45^\circ$. The bias voltage is set to be 25 V in this data set, and sample images with different bias voltages are shown in Figures \ref{fig:sim_mu_5V} and \ref{fig:sim_mu_100V}.


In Figure \ref{fig:sim_mu_25V} (a) to (c), muons are initialized with kinetic energy 100 keV, which, according to Figure \ref{fig:muon_energy_cali}, are not likely to penetrate through the CMOS without decaying into a high-energy electron. As result, what we see in these figures is one bright spot in the center, which corresponds to the muon depositing all its energy in a short range, and a long track of a high-energy minimum ionizing electron, which causes the large track probability. Also, although in all simulations the initial azimuth angle is fixed at $45^\circ$, in all 3 plots here with the same seed we see the track has azimuth angle around $135^\circ$, since this is the direction of the Michel electron (note it is a different direction from the muon direction, visible for higher energy panels in the figure). In this example, the 100 keV muon, after depositing all its energy, decays into a 40.5 MeV electron, a 14.6 MeV electron neutrino, and a 49.6 MeV muon antineutrino.

In Figure \ref{fig:sim_mu_25V} (d) to (f), muons are injected with kinetic energy 1 MeV, which has the largest ionizing loss as shown in Figure \ref{fig:muon_energy_cali}. As a result, images at this energy level show the brightest clusters of pixels in all simulation results. In Figure \ref{fig:sim_mu_25V} (d) where the zenith angle is smallest, we have the largest spot probability of the three plots. Also, in Figure \ref{fig:sim_mu_25V} (f), we see a narrower and less bright tail extending out of the main cluster, which causes the image to be classified with large worm probability. The Allpix$^2$ output shows that this is an electron from muon decay, which implies that, at the largest zenith angle, the muon travels farthest inside the CMOS so that it deposits all its energy and decays. In the examples we see here, at $\theta = 20^\circ$ the 1 MeV muon leaves CMOS with kinetic energy 0.743 MeV, and at $\theta = 50^\circ$ it leaves CMOS with kinetic energy 0.63 MeV. For $\theta = 80^\circ$, it deposits all 1 MeV of its initial kinetic energy and then decays into a 26.4 MeV electron, a 36.4 MeV electron neutrino, and a 41.7 MeV anti muon neutrino. 

In Figure \ref{fig:sim_mu_25V} (g) to (o), muons are injected with high enough energy that they are minimum ionizing particles (see Figure \ref{fig:muon_energy_cali}). They are very likely to penetrate through the CMOS without decaying or scattering. As a result, they typically have large track probability. An exception is Figure \ref{fig:sim_mu_25V} (l), in which a delta electron is ejected by the muon, which makes it a curved worm.

In addition to $-25$~V bias voltage, we produced simulation using bias voltage -5 V and -100 V, as shown in Figure~\ref{fig:sim_mu_5V} and Figure~\ref{fig:sim_mu_100V}. In each image set, in order to focus only on the effect of bias voltage, the random seeds we use for each image are exactly the same as the seeds used for Figure \ref{fig:sim_mu_25V} (j) to (o). As a result, Figure \ref{fig:sim_mu_25V} (l), \ref{fig:sim_mu_5V} (c), and \ref{fig:sim_mu_100V} (c) represent the same event under different bias voltage setups.

With larger bias voltage, electrons generated from ionizing radiation drift with greater vertical velocity, which leads to less charge diffusion visible in the image plane. Comparing Figure \ref{fig:sim_mu_25V} (l), \ref{fig:sim_mu_5V} (c), and \ref{fig:sim_mu_100V} (c), we see that the same event produces narrower signal deposition with larger bias voltage. Also, smaller spread in the pixel plane makes pixel clusters more track-like. Comparing \ref{fig:sim_mu_25V} (j), \ref{fig:sim_mu_5V} (a), and \ref{fig:sim_mu_100V} (a), we see that the track probability increases with larger bias voltage. As is evident in the images, the choice of bias voltage greatly impacts our simulation results and will be adjusted based on Monte Carlo simulations in section \ref{sec:dist_comp}.

\begin{figure*}[h!]
\begin{subfigure}{0.3333\textwidth}
  \centering
  \includegraphics[scale = 0.3]{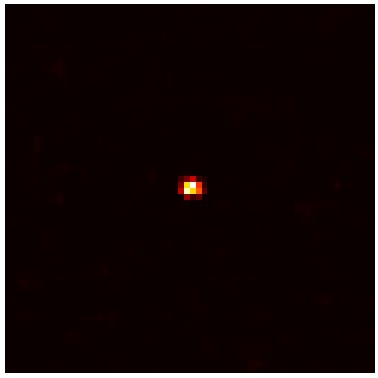}
  \captionsetup{justification=centering}
  \caption{10 keV, $\theta$ = $20^\circ$, T = 0,\\ W = 0, S = 1}
\end{subfigure}
\begin{subfigure}{0.3333\textwidth}
  \centering
  \includegraphics[scale = 0.3]{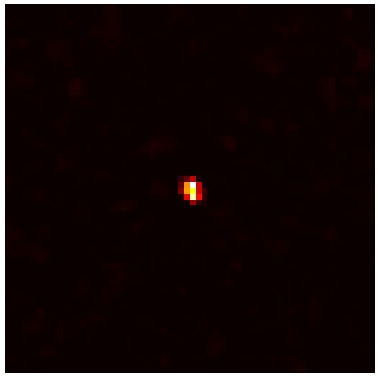}
  \captionsetup{justification=centering}
  \caption{10 keV, $\theta$ = $50^\circ$, T = 0,\\ W = 0, S = 1}
\end{subfigure}
\begin{subfigure}{0.3333\textwidth}
  \centering
  \includegraphics[scale = 0.3]{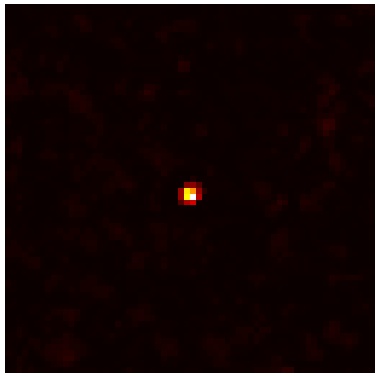}
  \captionsetup{justification=centering}
  \caption{10 keV, $\theta$ = $80^\circ$, T = 0,\\ W = 0, S = 1}
\end{subfigure}

\begin{subfigure}{0.3333\textwidth}
  \centering
  \includegraphics[scale = 0.3]{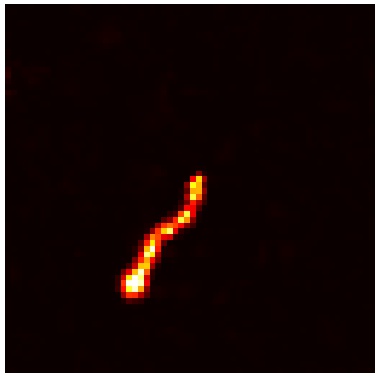}
  \captionsetup{justification=centering}
  \caption{100 keV, $\theta$ = $20^\circ$, T = 0.284,\\ W = 0.709, S = 0.006}
\end{subfigure}
\begin{subfigure}{0.3333\textwidth}
  \centering
  \includegraphics[scale = 0.3]{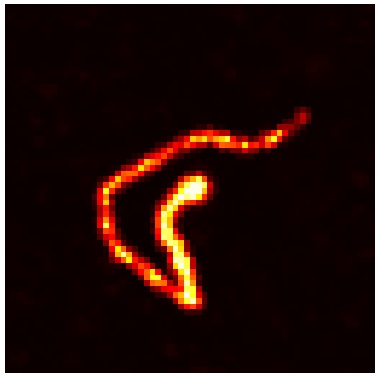}
  \captionsetup{justification=centering}
  \caption{100 keV, $\theta$ = $50^\circ$, T = 0,\\ W = 1, S = 0}
\end{subfigure}
\begin{subfigure}{0.3333\textwidth}
  \centering
  \includegraphics[scale = 0.3]{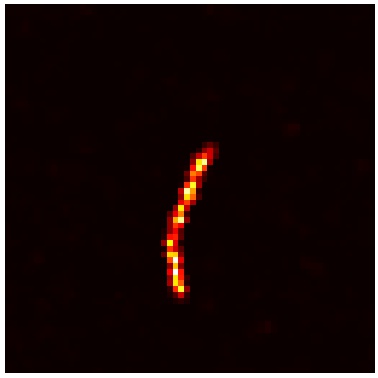}
  \captionsetup{justification=centering}
  \caption{100 keV, $\theta$ = $80^\circ$, T = 0.116,\\ W = 0.884, S = 0}
\end{subfigure}

\begin{subfigure}{0.3333\textwidth}
  \centering
  \includegraphics[scale = 0.3]{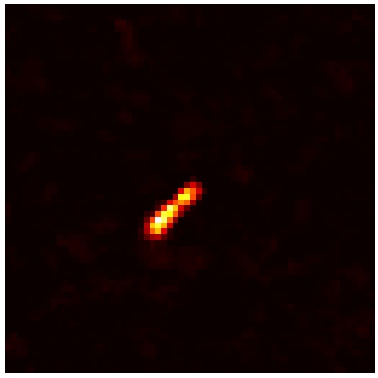}
  \captionsetup{justification=centering}
  \caption{1 MeV, $\theta$ = $20^\circ$, T = 0.931,\\ W = 0.069, S = 0}
\end{subfigure}
\begin{subfigure}{0.3333\textwidth}
  \centering
  \includegraphics[scale = 0.3]{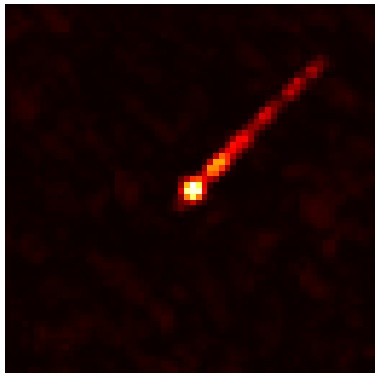}
  \captionsetup{justification=centering}
  \caption{1 MeV, $\theta$ = $50^\circ$, T = 0.985,\\ W = 0.015, S = 0}
\end{subfigure}
\begin{subfigure}{0.3333\textwidth}
  \centering
  \includegraphics[scale = 0.3]{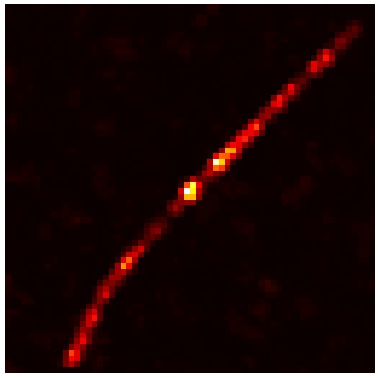}
  \captionsetup{justification=centering}
  \caption{1 MeV, $\theta$ = $80^\circ$, T = 1,\\ W = 0, S = 0}
\end{subfigure}

\begin{subfigure}{0.3333\textwidth}
  \centering
  \includegraphics[scale = 0.3]{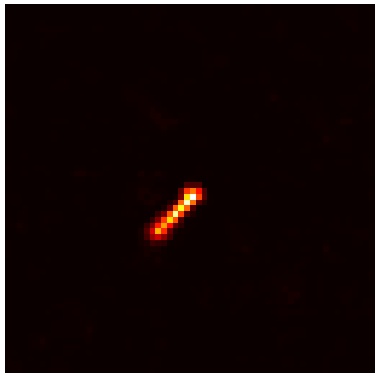}
  \captionsetup{justification=centering}
  \caption{10 MeV, $\theta$ = $20^\circ$, T = 0.987,\\ W = 0.013, S = 0}
\end{subfigure}
\begin{subfigure}{0.3333\textwidth}
  \centering
  \includegraphics[scale = 0.3]{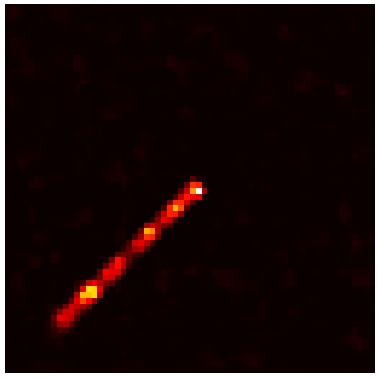}
  \captionsetup{justification=centering}
  \caption{10 MeV, $\theta$ = $50^\circ$, T = 0.998,\\ W = 0.002, S = 0}
\end{subfigure}
\begin{subfigure}{0.3333\textwidth}
  \centering
  \includegraphics[scale = 0.3]{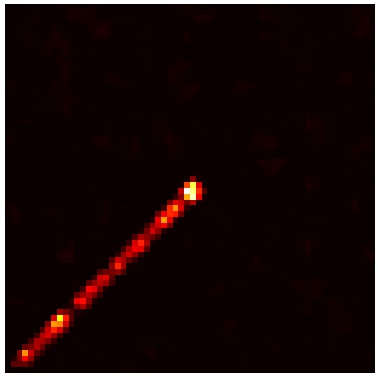}
  \captionsetup{justification=centering}
  \caption{10 MeV, $\theta$ = $80^\circ$, T = 0.999,\\ W = 0.001, S = 0}
\end{subfigure}

\begin{subfigure}{0.3333\textwidth}
  \centering
  \includegraphics[scale = 0.3]{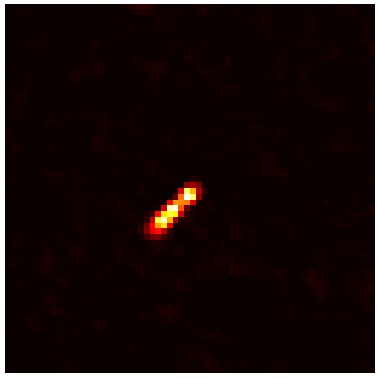}
  \captionsetup{justification=centering}
  \caption{100 MeV, $\theta$ = $20^\circ$, T = 0.969,\\ W = 0.031, S = 0}
\end{subfigure}
\begin{subfigure}{0.3333\textwidth}
  \centering
  \includegraphics[scale = 0.3]{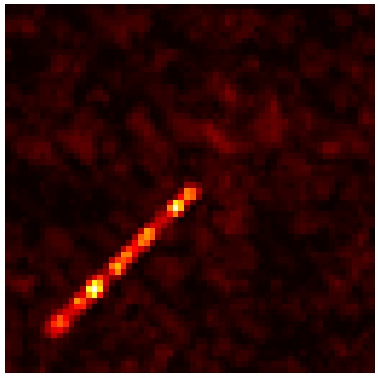}
  \captionsetup{justification=centering}
  \caption{100 MeV, $\theta$ = $50^\circ$, T = 0.991,\\ W = 0.008, S = 0}
\end{subfigure}
\begin{subfigure}{0.3333\textwidth}
  \centering
  \includegraphics[scale = 0.3]{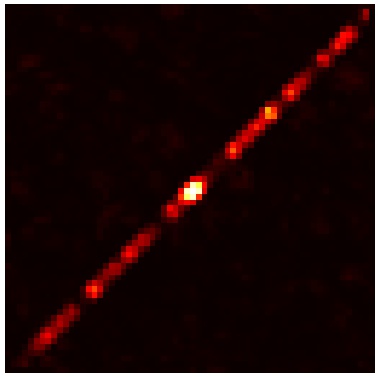}
  \captionsetup{justification=centering}
  \caption{100 MeV, $\theta$ = $80^\circ$, t = 0.999,\\ W = 0.001, S = 0}
\end{subfigure}

\caption{Example simulated electron events. The CNN classified probability is labeled as T (track), W (worm), and S (spot). Bias voltage is set to be -25 V, and $\phi$ is set to be $45^\circ$ in this simulation.}
\label{fig:sim_e_25V}
\end{figure*}

Analogous simulations are shown for electrons in Figure \ref{fig:sim_e_25V}. For electrons at initial kinetic energy 10 keV, according to Figure \ref{fig:electron_energy_cali}, they are not able to penetrate through the CMOS. As a result, electrons at this energy level deposit all their energy within a short range and stop, which produces a compact pixel cluster with large spot probability. 

At 100 keV, electrons have the largest ionizing energy loss in silicon as shown in Figure \ref{fig:electron_energy_cali}. They still at low enough energy to experience substnatial multiple scattering, which results in curved clusters of pixels and large worm probability. As shown in Figure~\ref{fig:sim_e_25V} (d) to (f), we see a lot of Coulomb scattering and large worm probability as expected.

In Figure~\ref{fig:sim_e_25V} (g) to (o), electrons with energy larger than 1 MeV are minimum ionizing particles (see Figure~\ref{fig:electron_energy_cali}). They are are very likely to penetrate through the CMOS without being deflected. As a result, they have large track probability as expected.

\begin{figure*}[h!]
\begin{subfigure}{0.3333\textwidth}
  \centering
  \includegraphics[scale = 0.3]{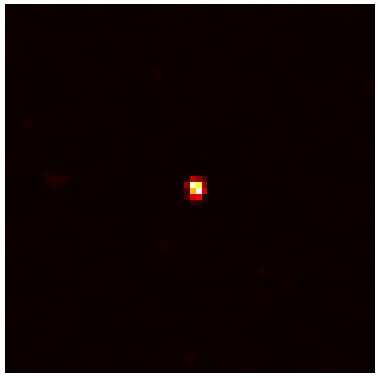}
  \captionsetup{justification=centering}
  \caption{10 keV, $\theta$ = $20^\circ$, T = 0,\\ W = 0, S = 1}
\end{subfigure}
\begin{subfigure}{0.3333\textwidth}
  \centering
  \includegraphics[scale = 0.3]{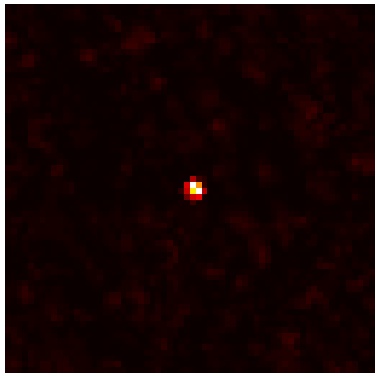}
  \captionsetup{justification=centering}
  \caption{10 keV, $\theta$ = $50^\circ$, T = 0,\\ W = 0, S = 1}
\end{subfigure}
\begin{subfigure}{0.3333\textwidth}
  \centering
  \includegraphics[scale = 0.3]{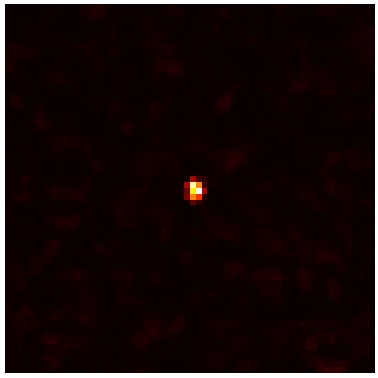}
  \captionsetup{justification=centering}
  \caption{10 keV, $\theta$ = $80^\circ$, T = 0,\\ W = 0, S = 1}
\end{subfigure}

\begin{subfigure}{0.3333\textwidth}
  \centering
  \includegraphics[scale = 0.3]{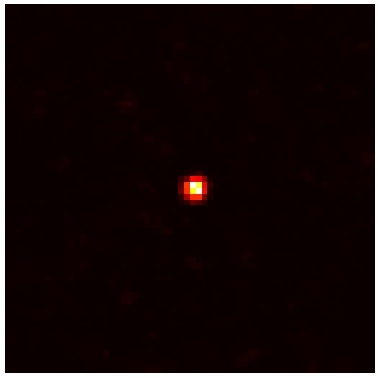}
  \captionsetup{justification=centering}
  \caption{100 keV, $\theta$ = $20^\circ$, T = 0,\\ W = 0, S = 1}
\end{subfigure}
\begin{subfigure}{0.3333\textwidth}
  \centering
  \includegraphics[scale = 0.3]{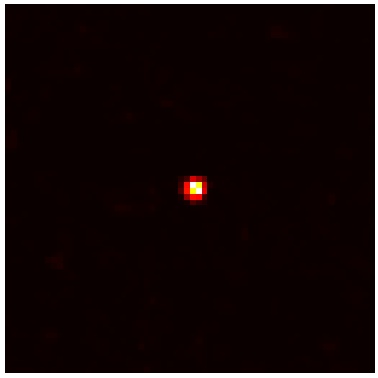}
  \captionsetup{justification=centering}
  \caption{100 keV, $\theta$ = $50^\circ$, T = 0,\\ W = 0, S = 1}
\end{subfigure}
\begin{subfigure}{0.3333\textwidth}
  \centering
  \includegraphics[scale = 0.3]{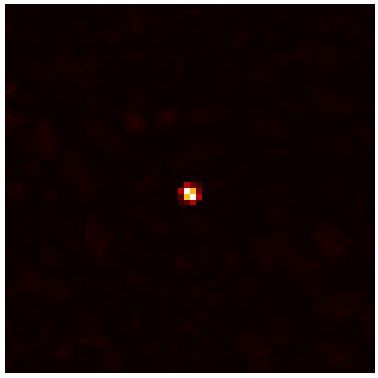}
  \captionsetup{justification=centering}
  \caption{100 keV, $\theta$ = $80^\circ$, T = 0,\\ W = 0, S = 1}
\end{subfigure}

\begin{subfigure}{0.3333\textwidth}
  \centering
  \includegraphics[scale = 0.3]{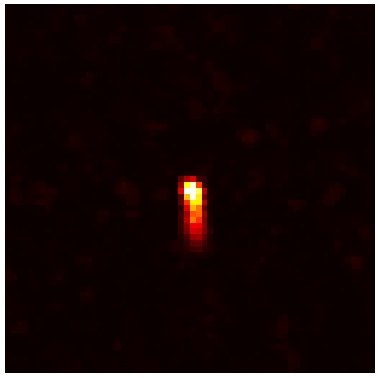}
  \captionsetup{justification=centering}
  \caption{1 MeV, $\theta$ = $20^\circ$, T = 0.671,\\ W = 0.301, S = 0.021}
\end{subfigure}
\begin{subfigure}{0.3333\textwidth}
  \centering
  \includegraphics[scale = 0.3]{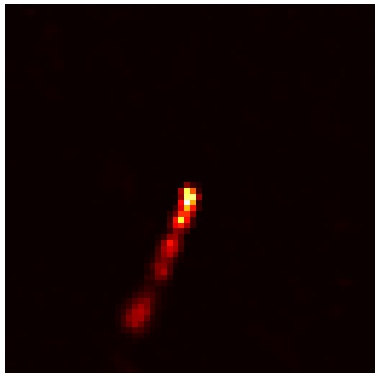}
  \captionsetup{justification=centering}
  \caption{1 MeV, $\theta$ = $50^\circ$, T = 0.911,\\ W = 0.089 S = 0}
\end{subfigure}
\begin{subfigure}{0.3333\textwidth}
  \centering
  \includegraphics[scale = 0.3]{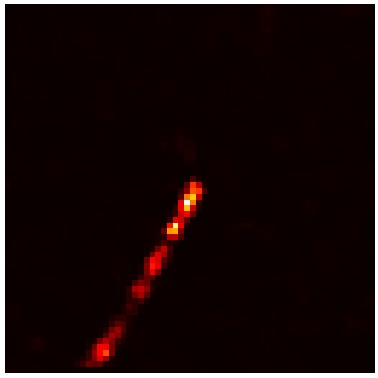}
  \captionsetup{justification=centering}
  \caption{1 MeV, $\theta$ = $80^\circ$, T = 0.996,\\ W = 0.004, S = 0}
\end{subfigure}

\begin{subfigure}{0.3333\textwidth}
  \centering
  \includegraphics[scale = 0.3]{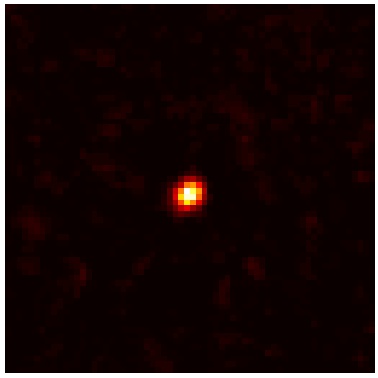}
  \captionsetup{justification=centering}
  \caption{10 MeV, $\theta$ = $20^\circ$, T = 0,\\ W = 0, S = 1}
\end{subfigure}
\begin{subfigure}{0.3333\textwidth}
  \centering
  \includegraphics[scale = 0.3]{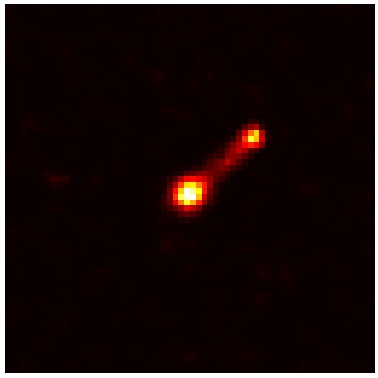}
  \captionsetup{justification=centering}
  \caption{10 MeV, $\theta$ = $50^\circ$, T = 0.708,\\ W = 0.292, S = 0}
\end{subfigure}
\begin{subfigure}{0.3333\textwidth}
  \centering
  \includegraphics[scale = 0.3]{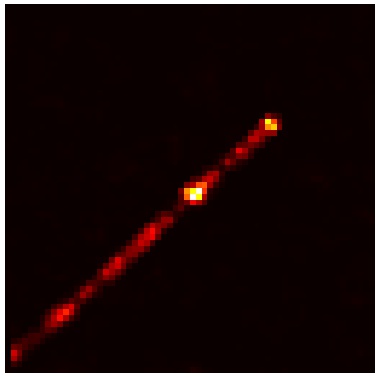}
  \captionsetup{justification=centering}
  \caption{10 MeV, $\theta$ = $80^\circ$, T = 0.999,\\ W = 0.001, S = 0}
\end{subfigure}

\begin{subfigure}{0.3333\textwidth}
  \centering
  \includegraphics[scale = 0.3]{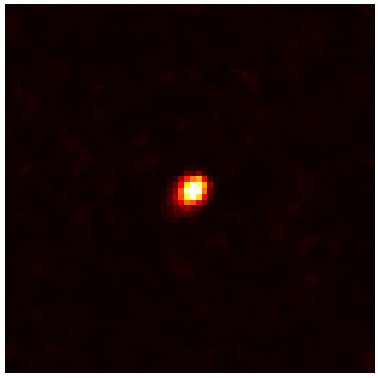}
  \captionsetup{justification=centering}
  \caption{100 MeV, $\theta$ = $20^\circ$, T = 0,\\ W = 0, S = 1}
\end{subfigure}
\begin{subfigure}{0.3333\textwidth}
  \centering
  \includegraphics[scale = 0.3]{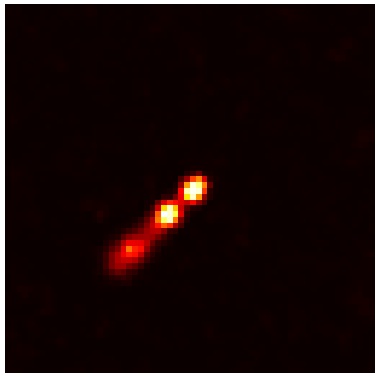}
  \captionsetup{justification=centering}
  \caption{100 MeV, $\theta$ = $50^\circ$, T = 0.955,\\ W = 0.045, S = 0}
\end{subfigure}
\begin{subfigure}{0.3333\textwidth}
  \centering
  \includegraphics[scale = 0.3]{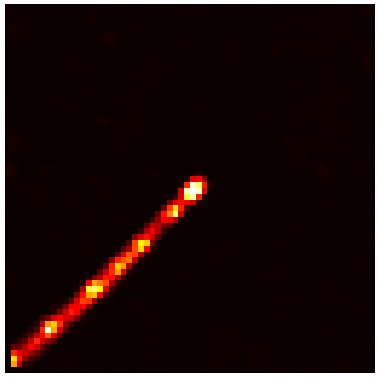}
  \captionsetup{justification=centering}
  \caption{100 MeV, $\theta$ = $80^\circ$, T = 0.995,\\ W = 0.005, S = 0}
\end{subfigure}

\caption{Example simulated photon events. The CNN classified probability is labeled as T (track), W (worm), and S (spot). Bias voltage is set to be -25 V, and $\phi$ is set to be $45^\circ$ in this simulation.}
\label{fig:sim_photon_25V}
\end{figure*}

For photons, simulated events are shown in Figure \ref{fig:sim_photon_25V}. As shown in Figure \ref{fig:gamma_energy_cali}, at 10 keV the dominating photon interaction is the photoelectric effect, which produces an electron with nearly the same energy as the original photon. The $sim$10 keV electron then produces a spot-like image just as in Figure \ref{fig:sim_e_25V}. From 100 keV to 10 MeV, Compton scattering is the dominant interaction type and the expected electron energy increases with photon energy. As a result, at 100 keV the low-energy electron produced by Compton scattering leaves spot-like images, and as photon energy increases the outgoing electron produces a more and more track-like image. At 100 MeV, pair production dominates, producing one electron and one positron with total energy 100 MeV. These high energy electrons are likely to leave long and straight trajectories, and they are also likely to have initial direction nearly colinear. As a result, after charge dispersion in the CMOS, their individual trajectories are indistinguishable in the final processed image.

Comparing simulated image examples of muons (Figure~\ref{fig:sim_mu_25V}), electrons (Figure~\ref{fig:sim_e_25V}), and photons (Figure~\ref{fig:sim_photon_25V}) with examples of experimental images collected by DECO (figure \ref{fig:real_image}), we see that our simulation produces results similar to experimental data in many aspects. However, due to the stochastic nature of particle production and energy loss, it is difficult to draw quantitative conclusions from theses examples.  We perform more quantitative data - Monte Carlo comparison in Section \ref{sec:dist_comp}.

\subsection{Probability Distribution Classified by the CNN}
\label{sec:prob_dist}

In this section, we focus on the relationship between particle energy and event type as classified by the CNN. For muons, electrons, and photons separately, at each of the 8 energy levels selected uniformly in log-space from 10 keV to 100 GeV, a Monte-Carlo simulation is performed by randomly generating incident particles according to their specific angular distribution. All simulated events are fed into the CNN to be classified. Any event with $P_{track, worm, spot, or noise} \geq 50\%$ is given the corresponding label; otherwise it is labeled as ``other''. The rates of events classified as track, worm, spot, noise, and other are plotted as a function of energy for each particle type.

\begin{figure}[h!]
\centering
  \includegraphics[width=0.5\textwidth]{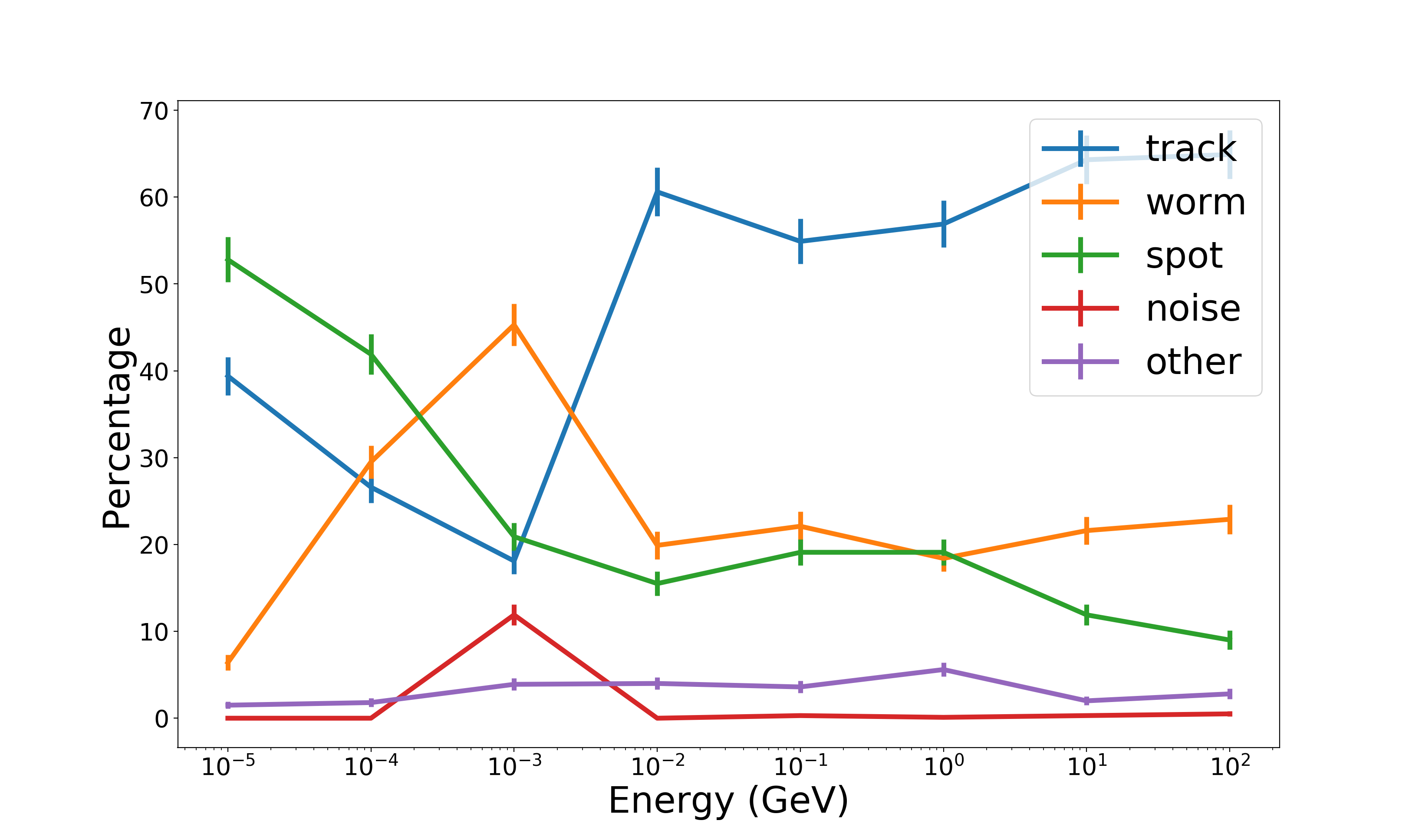}
  \caption{Percentage of Monte Carlo simulated muon events at different energy levels classified as track, worm, spot, noise, or other by the CNN}
  \label{fig:muon_confusion}
\end{figure}

Since most muon events DECO collects are atmospheric muons produced by cosmic-ray interactions, muons are simulated according to the procedure in section \ref{sec:MC_sim} with fixed energy levels and randomly generated zenith angles according to Equation \ref{eq:muon_flux}. The results are shown in Figure \ref{fig:muon_confusion}.

At 10 keV, most muons are classified as spot or track. Since a 10 keV muon deposits all its energy without going far inside the CMOS, it creates a bright point-like cluster of pixels, as shown in the center of figure \ref{fig:sim_e_25V} (a) (b) (c). Then the muon decay produces a minimum ionizing electron which leaves a less bright, long tail. As a result, it might be regarded as either a spot or track by the CNN depending on the electron direction and its deposited energy.

At increasing muon energy, the spot and track rates decrease while the worm rate increases. This is explained by the example in Figure \ref{fig:sim_mu_25V} (f), where a muon experiences large ionizing loss, deposits all it energy, and decays into an electron in a random direction. Compared to the 10 keV case, at these energies muons are able to penetrate greater distance inside the CMOS; however, if it still loses all its energy inside the CMOS, a minimum ionizing electron will be produced in a random direction, which makes the image a curved worm. Large ionization loss of muons at this energy level might explain the high rate of noise events at 1 MeV as well.

With energy 10 MeV and greater, muons are minimum ionizing particles inside silicon. Muons are less likely to decay or be deflected, so the track rate is high and varies little with muon energy. Delta rays in a fraction of events (as shown for example in Figure \ref{fig:sim_mu_25V} (l)) cause a moderately high worm rate even among these minimum-ionizing muons typically produced by cosmic rays.

       

\begin{figure}[h!]
\centering
  \includegraphics[width=0.5\textwidth]{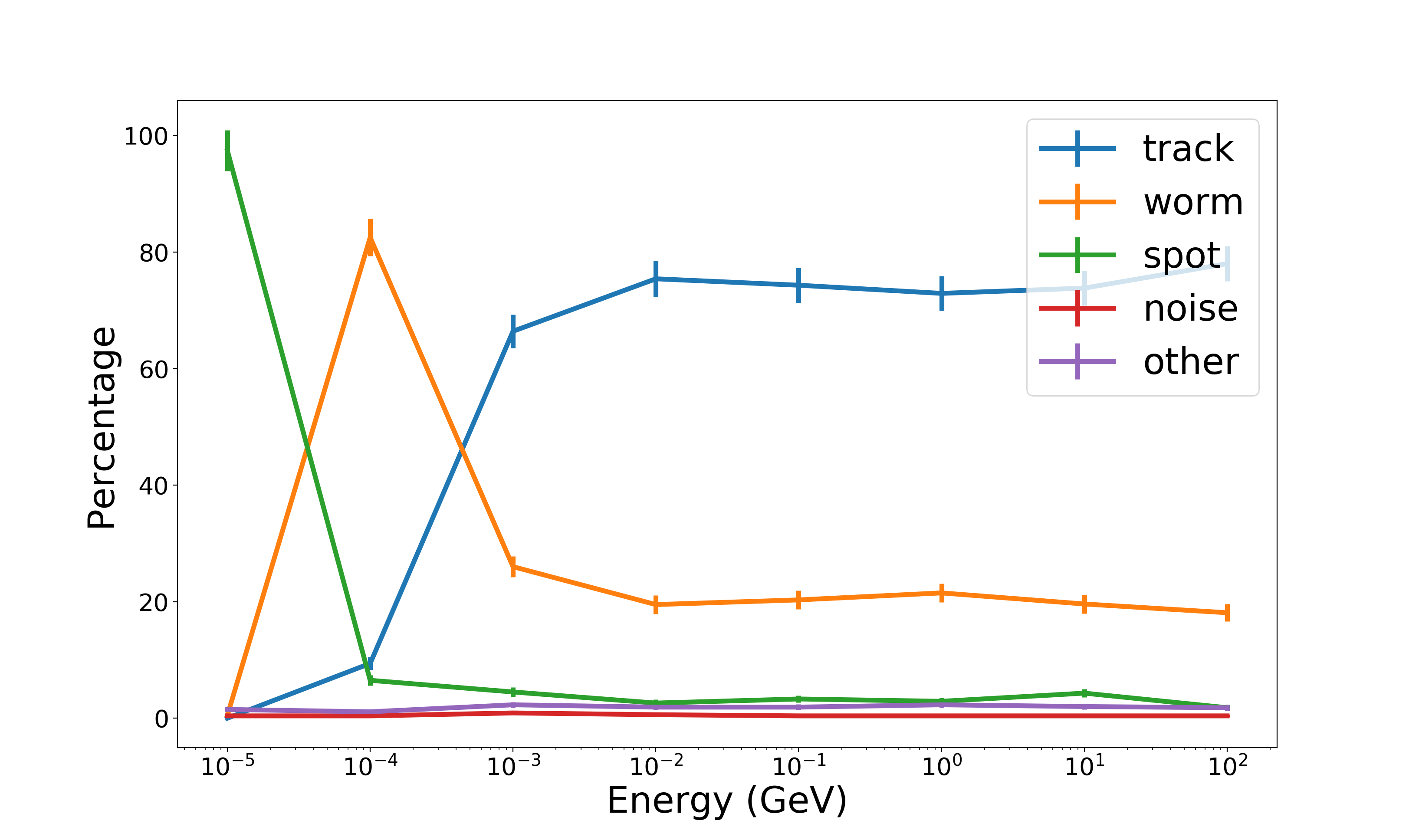}
  \caption{Percentage of Monte Carlo simulated electron events at different energy levels classified as tracks, worms, spots, noises, or others by the CNN}
  \label{fig:electron_confusion}
\end{figure}

For electrons, we assume an isotropic angular distribution since many of the electron events in DECO come from radioactive background, e.g. beta emitting isotopes in the cell phone material or air surrounding the sensor. The weight function $\cos(\theta) * \frac{1 - \cos(\theta)}{2}$ is used, in which the first $\cos(\theta)$ represents the detector effective area and the rest, along with uniformly selected $\phi$, is constructed to uniformly sample points on the sphere.

At 10 keV, almost all electron events are classified as spot. As seen in Figure \ref{fig:sim_e_25V} (a) (b) (c), at this low energy electrons quickly desposit all their energy inside the sensor. As a result, the trajectory is very short. At 100 keV, as shown in Figure~\ref{fig:sim_e_25V} (d) (e) (f), electrons are subject to large Coulomb scattering, so the worm rate is high. At 1 MeV and beyond, electrons are in the minimum ionizing energy regime with low Coulomb scattering, so most of the events are classified as tracks with negligible dependence on energy. As for muon events, a substantial fraction ($sim$20\%) are classified as worms likely due to delta electrons.

\begin{figure}[h!]
\centering
  \includegraphics[width=0.5\textwidth]{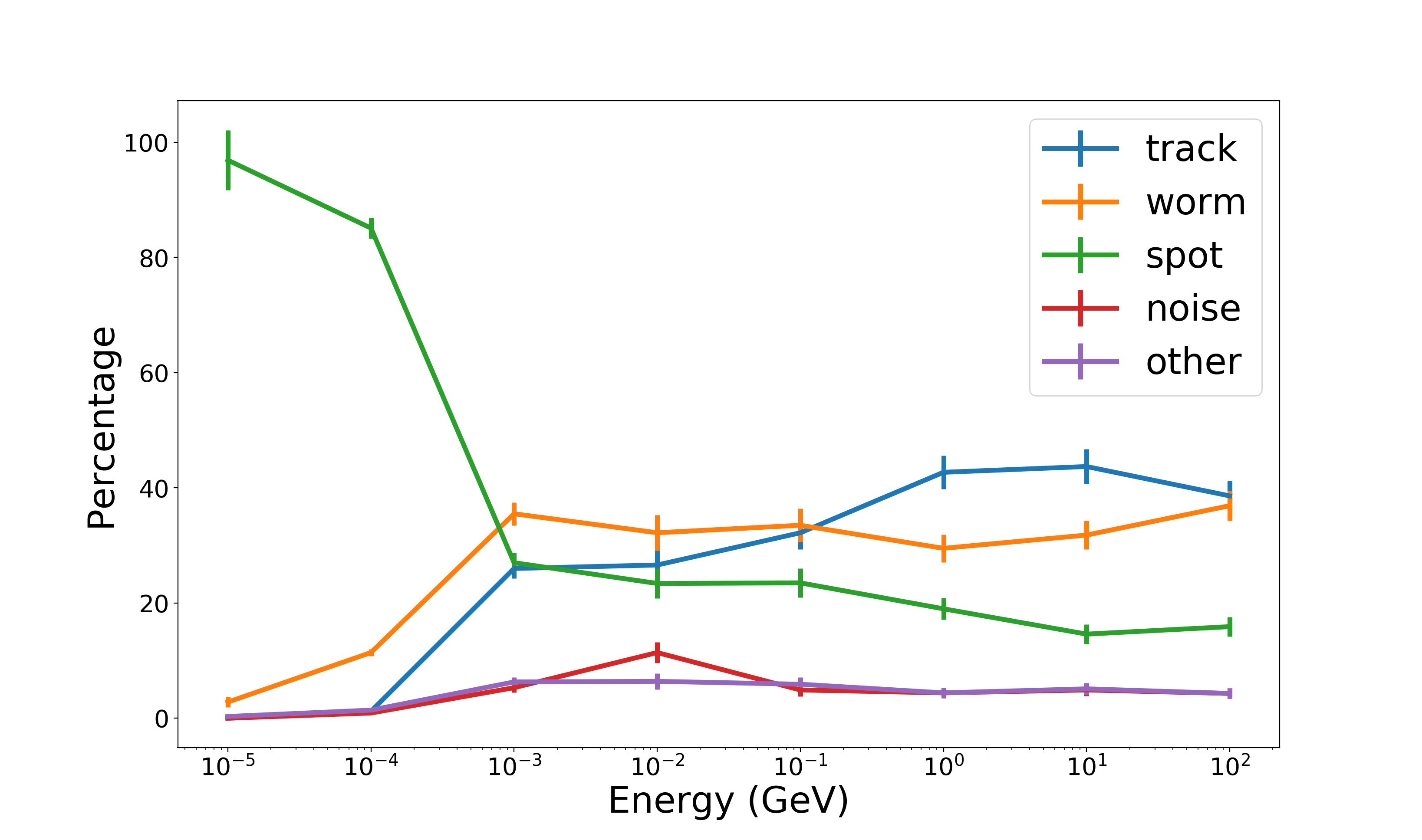}
  \caption{Percentage of Monte Carlo simulated photon events at different energy levels classified as track, worm, spot, noise, or other by the CNN}
  \label{fig:photon_confusion}
\end{figure}

For photons, the same angular distribution as electrons is used, and the result is show in Figure~\ref{fig:photon_confusion}. As shown in Figure~\ref{fig:sim_photon_25V}, at energy below $sim$0.3~MeV, the low energy electron produced by the photoelectric effect or Compton scattering is likely to be absorbed quickly which results in high probability of being classified as spot. At higher energy, although photons produce high energy electrons which are expected to show similar behavior as the electron classification plot in Figure~\ref{fig:electron_confusion}, we see that photons have a similar track and worm rate even at GeV energies, while high energy electrons in Figure~\ref{fig:electron_confusion} have a track rate around 3 times larger than their worm rate. This difference is explained by the fact that, in photon events, positions where electrons are produced by photon interaction is randomly distributed inside the bulk of the CMOS, but for electron events all electrons begin interacting at the center of the CMOS top surface. Compared to electrons initialized on the top surface, electrons produced by photon interactions inside the CMOS create electron-hole pairs which experience larger diffusion while drifting to the surface. This can explain the greater rate of track classification seen for electrons than photons in this high energy range.

\subsection{Data – Monte Carlo comparison}
\label{sec:dist_comp}

In this section we perform a quantitative analysis of parameters calculated from Monte Carlo simulated images, including their similarity to experimental data and their dependence on bias voltage.  One challenge associated with such simulations is that, different from atmospheric muons whose flux can be well described by equation \ref{eq:muon_flux}, the flux of electrons and photons is background-dependent and can be difficult to approximate. In order to simplify this comparison we select data images and simulated muon images classified by the CNN with $P_{track} > 50\%$.  Muons are randomly generated according to the procedure in section \ref{sec:MC_sim}, their images are classified by the CNN, and all the simulated images classified by the CNN with $P_{track} > 50\%$ are collected. After image cleaning following the procedure described in Appendix \ref{app:image_clean}, their image parameters are extracted and plotted against those of data events classified by the CNN with $P_{track} > 50\%$ and after image cleaning.

\begin{figure}[h!]
\begin{subfigure}{0.5\textwidth}
  \centering
  \includegraphics[scale = 0.23]{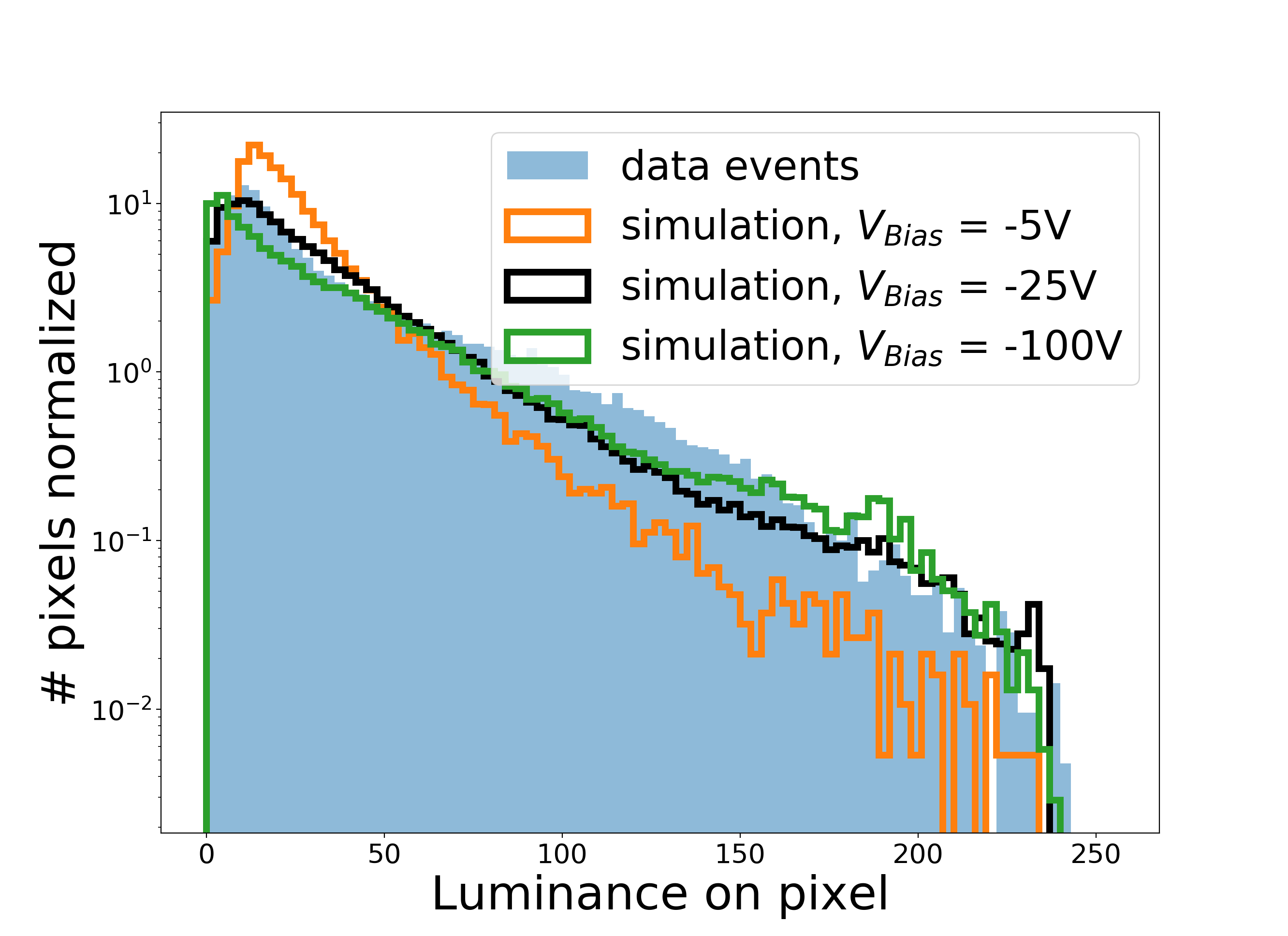}
  \caption{Distribution of luminance on image pixels.}
\end{subfigure}
\begin{subfigure}{0.5\textwidth}
  \centering
  \includegraphics[scale = 0.23]{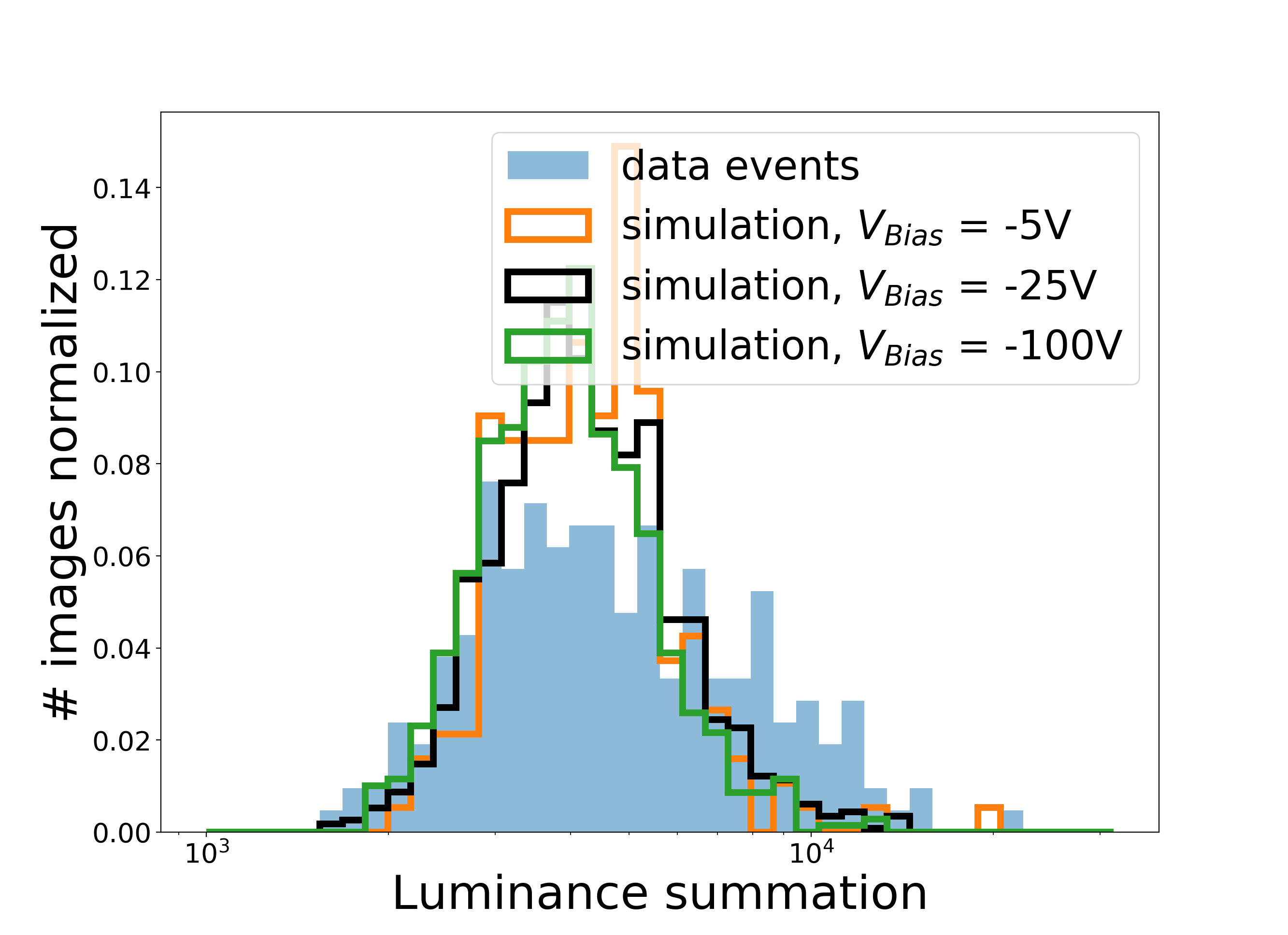}
  \caption{Distribution of summed luminance on image.}
\end{subfigure}

\begin{subfigure}{0.5\textwidth}
  \centering
  \includegraphics[scale = 0.23]{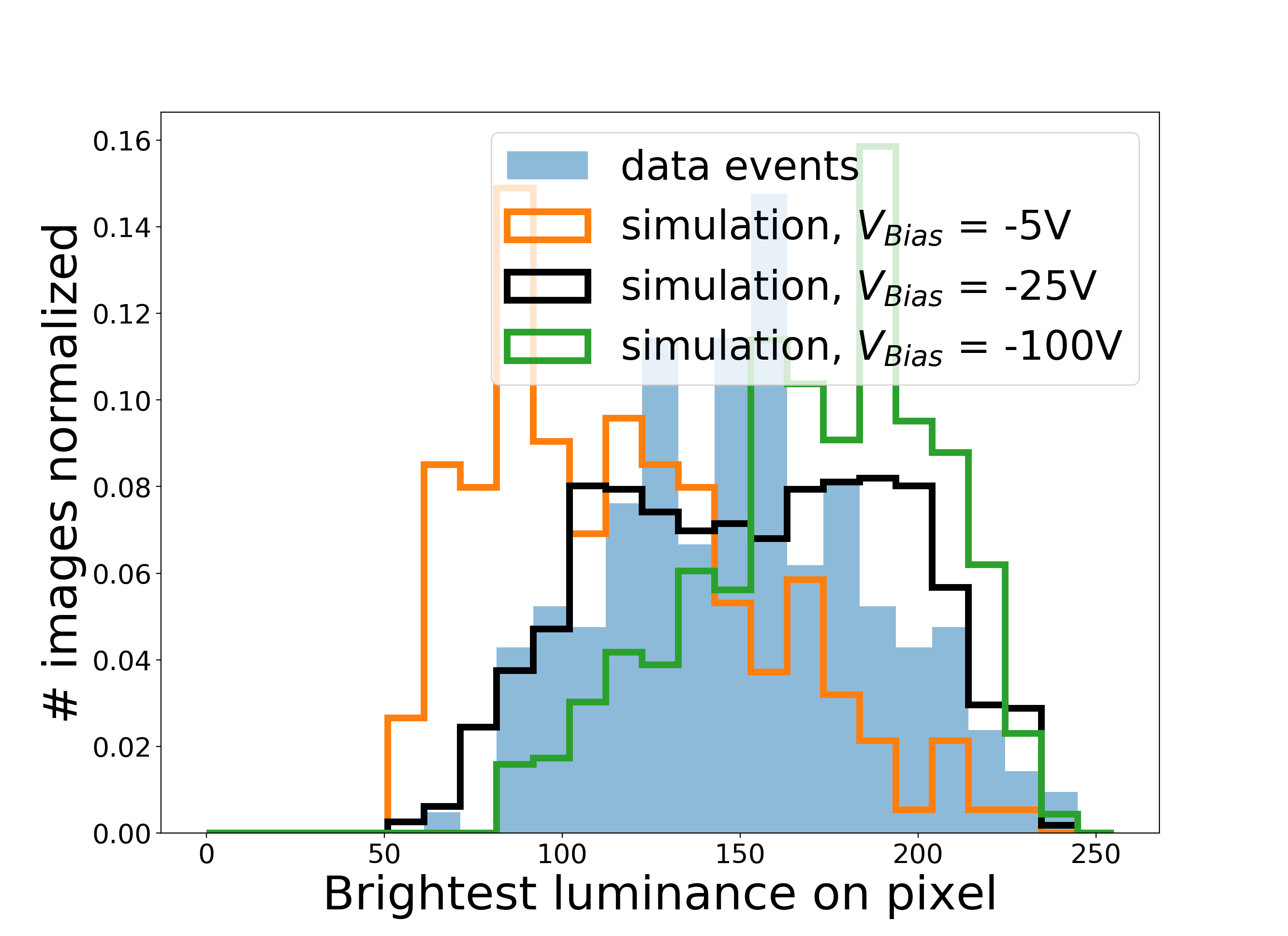}
  \caption{Distribution of brightest pixel luminance on image.}
\end{subfigure}

\caption{Comparison of luminance distribution between simulated events at different bias voltages with $P_{track} \geq 50\%$ and track events in data with $P_{track} \geq 50\%$.}
\label{fig:Luminance_check}
\end{figure}

In Figure~\ref{fig:Luminance_check} (a), the distribution of luminance for every pixel of every image is plotted. Since our colored images have 8 bits in each pixel, the range of this distribution is 0 to 255. As bias voltage increases, electrons generated from ionizing radiation drift with larger vertical velocity, which results in less lateral charge diffusion visible in the image plane. Since the total amount of charge produced by energy loss remains constant, the slope in Figure~\ref{fig:Luminance_check} (a) decreases as smaller bias voltages increase charge diffusion. Although both -25 V and -100 V give us a close match with experimental data images, there is still some discrepancy at luminance 100 - 150, which could be caused by our simplified model of electric field, image processing, and muon-only simulation.

In Figure \ref{fig:Luminance_check} (b), the luminance on each image is summed over pixels and the distribution of this total luminance is plotted. Since total luminance is directly related to the energy loss, which is independent of charge diffusion, distributions with different bias voltages are similar. However, the QDC we use, which affects the conversion ratio between charge and signal, does affect this distribution, and the reasonable agreement we see here between simulated events and experimental data events indicates that our current QDC simulation works as intended.

In Figure~\ref{fig:Luminance_check} (c), the distribution of luminance at the brightest pixel in each image is plotted. Since large bias voltages decreases the charge diffusion, the charge is more concentrated, increasing the expected brightest luminance. As a result, larger bias voltages shift the plot to the right. Here we see that -25 V bias voltage gives us the best match between simulation ane experimental data.

\begin{figure}[h!]
\begin{subfigure}{0.5\textwidth}
  \centering
  \includegraphics[scale = 0.23]{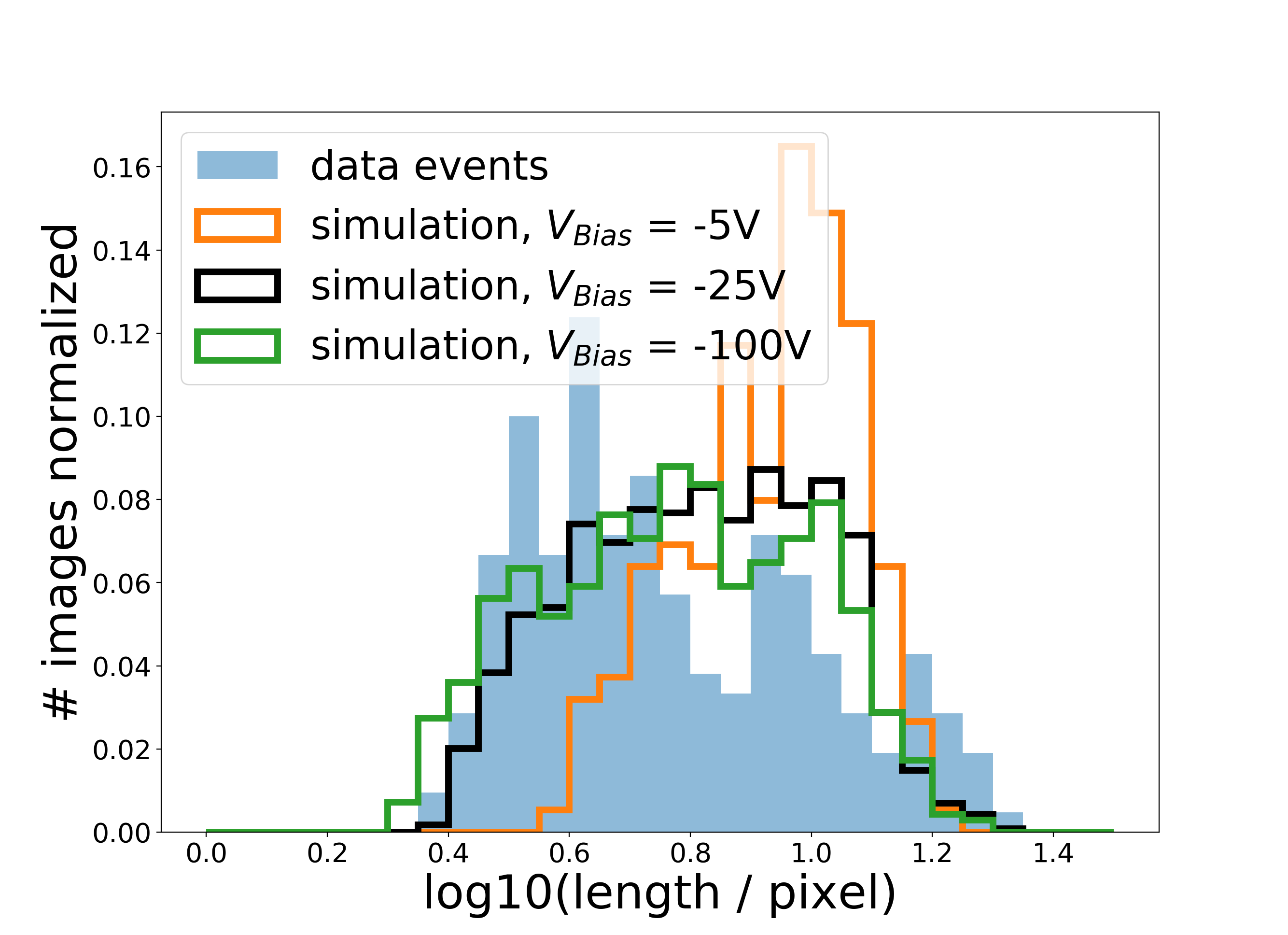}
  \caption{Distribution of Hillas length.}
\end{subfigure}
\begin{subfigure}{0.5\textwidth}
  \centering
  \includegraphics[scale = 0.23]{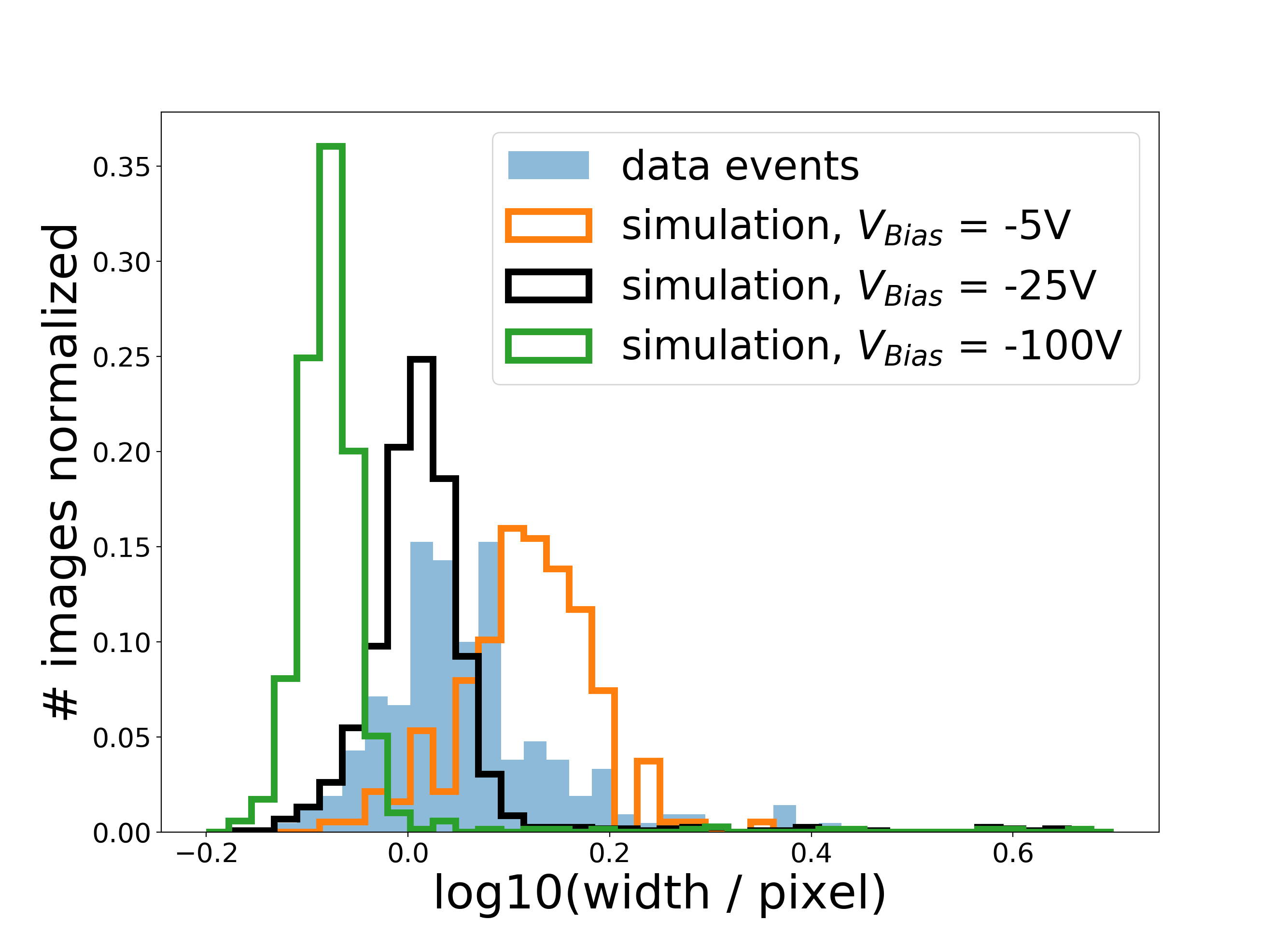}
  \caption{Distribution of Hillas width.}
\end{subfigure}

\caption{Distribution of Hillas parameters \cite{Hillas_calculation} between simulated events at different bias voltages with $P_{track} \geq 50\%$ and track events in data with $P_{track} \geq 50\%$.}
\label{fig:Hillas_check}
\end{figure}

In addition to the luminance distribution, Hillas parameters, which parameterize images as described in \cite{Hillas_calculation}, are calculated for each image.  The distribution of Hillas width and length is plotted in Figure~\ref{fig:Hillas_check}. This technique for examining a deposited charge image on a photosensor is typically used for imaging atmospheric Cherenkov telescopes. For Hillas length, since we only select events with $P_{track} > 50\%$ to make the histogram, these example images tend to be long straight lines whose length is largely determined by the zenith angle instead of charge diffusion. As a result, different bias voltages does not make a significant difference in Hillas length distribution unless the voltage is particularly low. On the other hand, the width of track events is largely determined by lateral charge diffusion, so larger bias voltages make smaller Hillas width distributions. The best match still appears at -25 V.

\begin{figure}[h!]
\begin{subfigure}{0.5\textwidth}
  \centering
  \includegraphics[scale = 0.23]{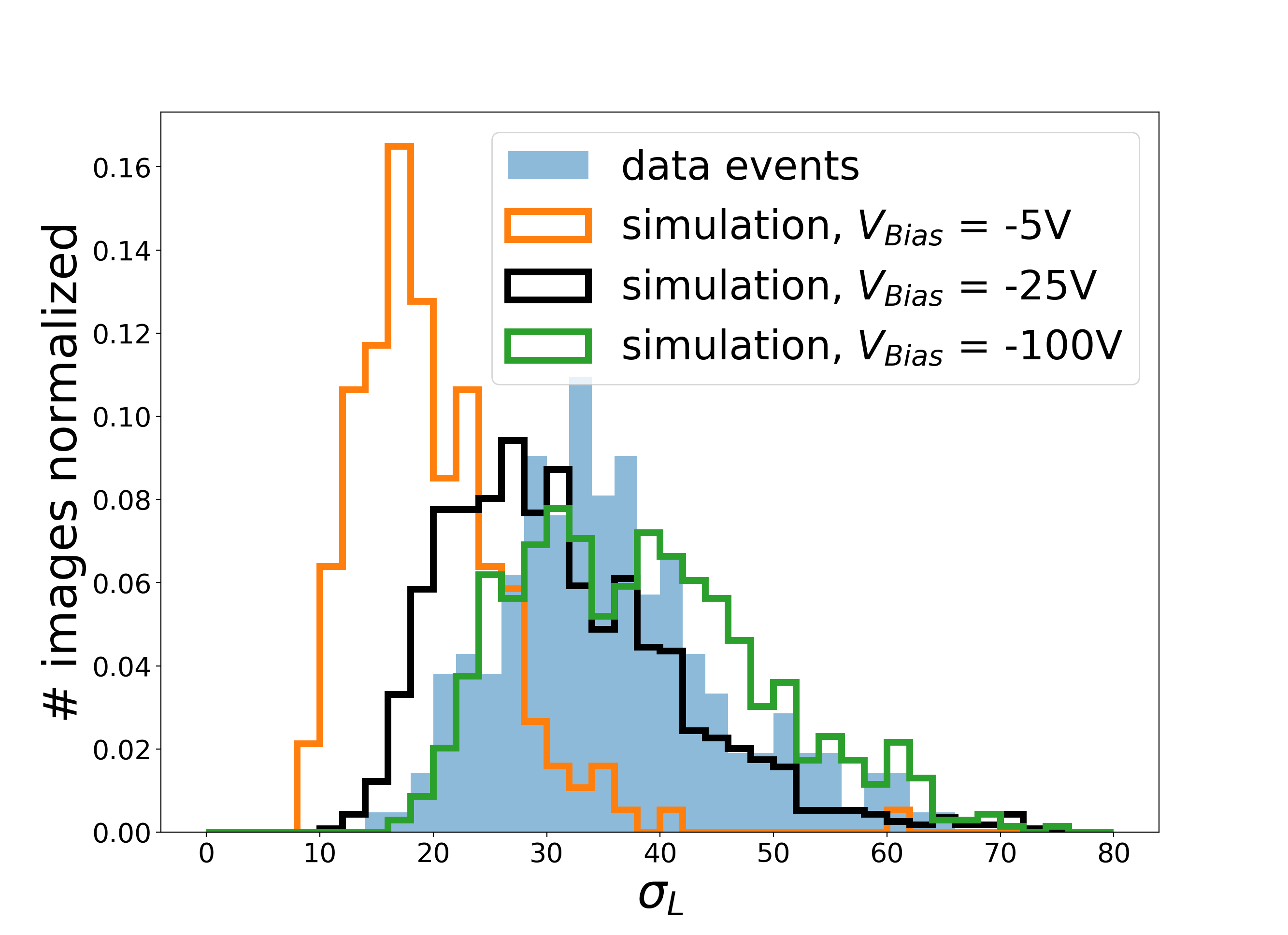}
  \caption{Distribution of $\sigma_{L}$.}
\end{subfigure}
\begin{subfigure}{0.5\textwidth}
  \centering
  \includegraphics[scale = 0.23]{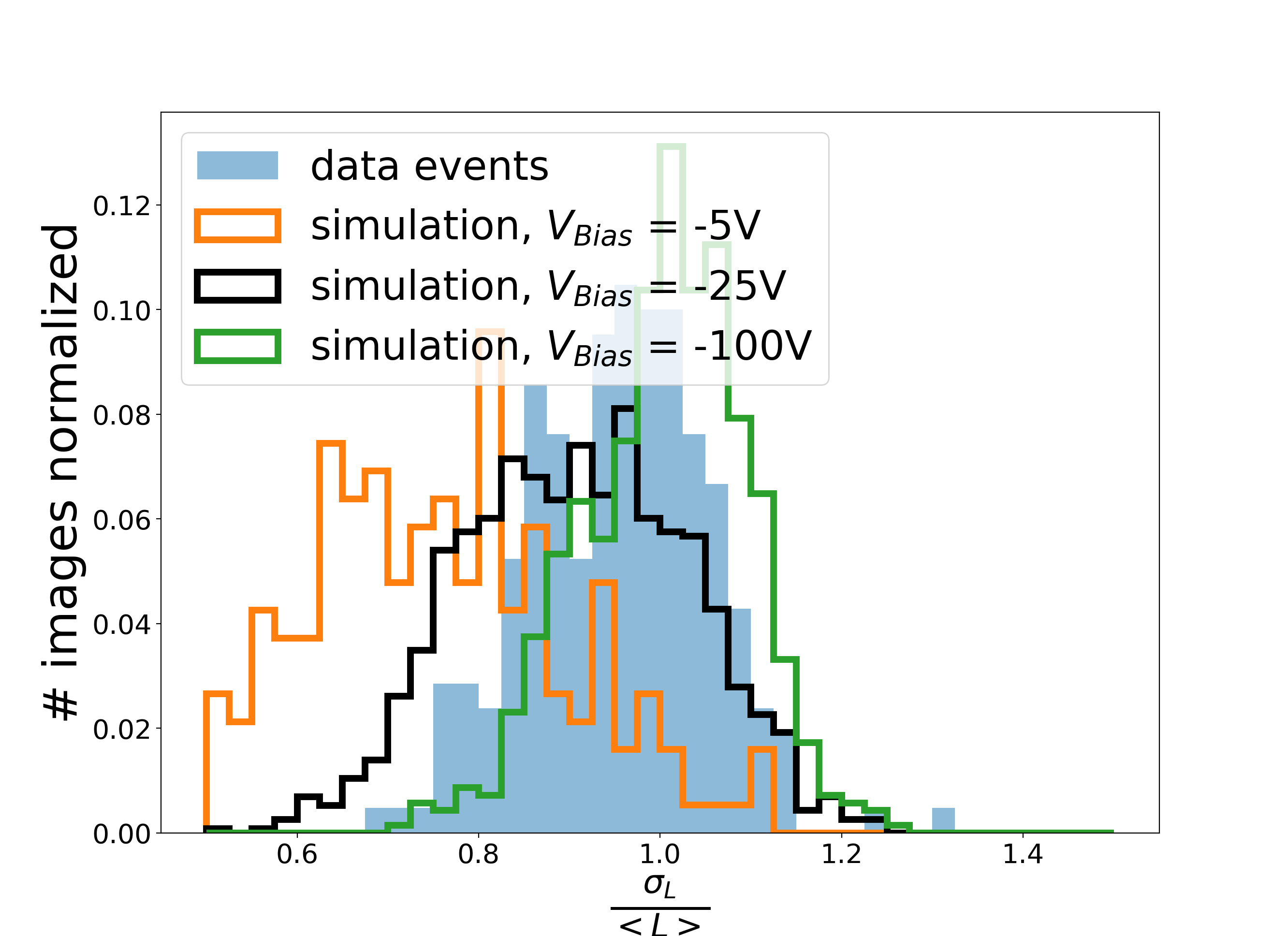}
  \caption{Distribution of $\frac{\sigma_{L}}{\langle L\rangle}$.}
\end{subfigure}

\caption{Distribution of $\sigma_{L}$ and $\frac{\sigma_{L}}{\langle L\rangle}$ between simulated events at different bias voltages with $P_{track} \geq 50\%$ and track events in data with $P_{track} \geq 50\%$.}
\label{fig:Moment_check}
\end{figure}

Next, for every image the standard deviation of luminance among pixels ($\sigma_L$) and standard deviation over mean $\frac{\sigma_L}{\langle L\rangle}$ are calculated.  Their distributions are plotted in Figure~\ref{fig:Moment_check}. As bias voltage increases, smaller charge diffusion causes larger standard deviation of luminance distribution, which shifts Figure~\ref{fig:Moment_check} (a) to the right. However, larger bias voltage also makes mean luminance larger.  Here we see that both -25 V and -100 V give us a close match between Monte Carlo and experimental data.


\section{Conclusion and Outlook}

In order to understand the underlying physics behind images recorded by DECO and other similar cell phone projects, we used Allpix$^2$ to simulate the interaction of cosmic-ray particles and particles from radioactive decay in cell phone CMOS chips. We used CMOS model parameters similar to those for the HTC Wildfire phone model, and similar image processing procedures, including Bayer filtering, white balancing.  We included real background noise by summing samples from real experimental data images recorded in dark conditions. The simulation results show physical properties similar to those observed in collected DECO data, including deposited energy and event morphology distribution, and the data-Monte Carlo comparison shows good agreement after selecting the bias voltage that results in the best agreement.  For three particle types spanning eight decades in incident energy, we validated our Monte Carlo simulations by comparing to analytical expectations of energy loss and interaction probability.  Example images produced from simulations at a range of energies, particle types, and incident directions compare well to experimental data, as do quantitative metrics calculated from both experimental and simulated data.


Future work includes (1) using RAW camera image data in order to be less sensitive to post-processing performed by camera hardware and software (2) using Monte Carlo simulation with known incident particle truth, rather than human determined labels, to train the CNN; (3) developing algorithms for energy and direction reconstruction as well as particle identification, which can be validated on our Monte Carlo simulation and applied to experimental data (4) generalizing the simulations and comparison to data to a larger variety of device architecture.

\section*{Appendix: Image Cleaning Algorithm}
\label{app:image_clean}

\begin{figure}[h!]
\centering
  \includegraphics[width=0.4\textwidth]{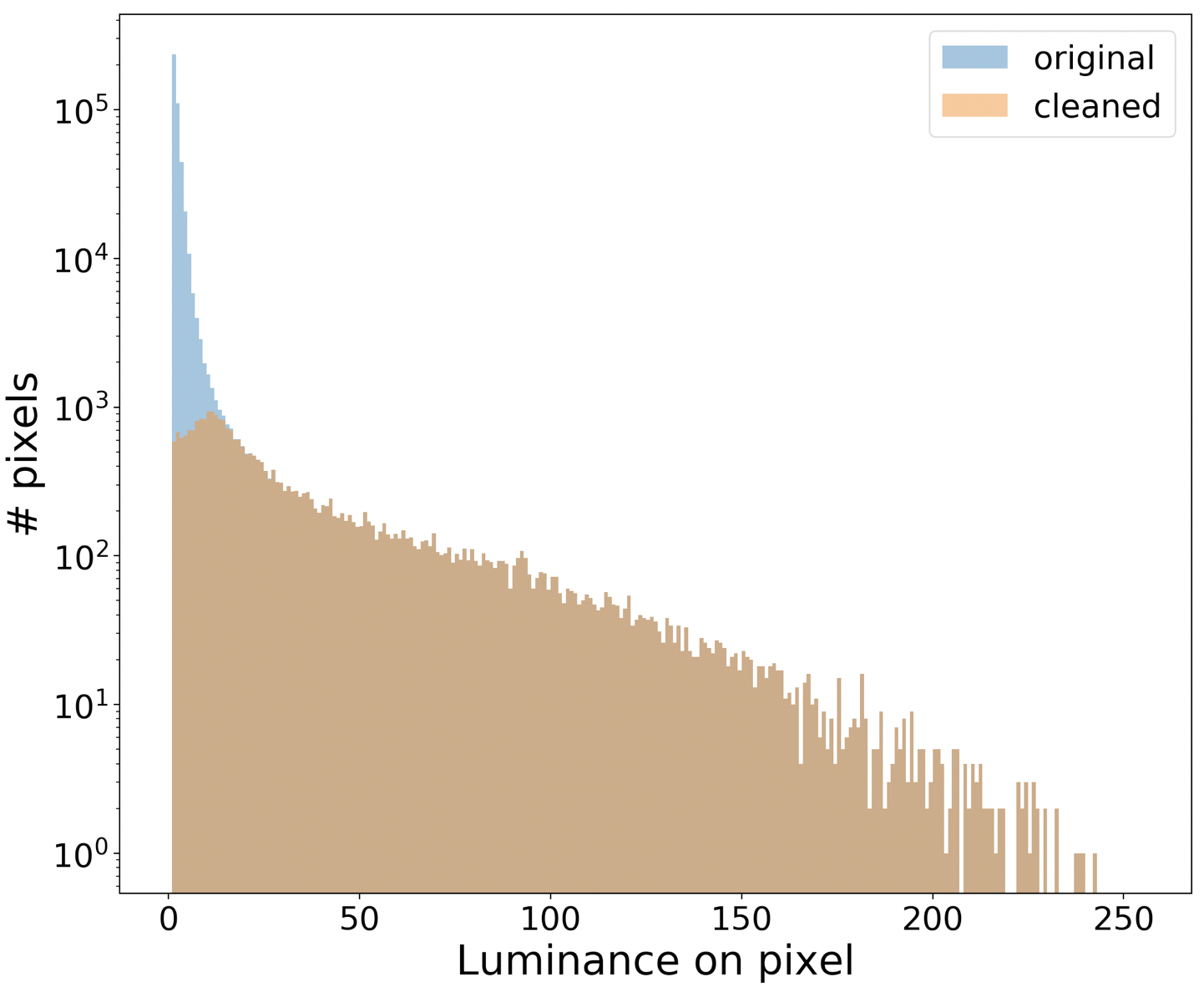}
  \caption{Luminance distribution of track events in experimental data, with and without image cleaning. Threshold value is set to be 10.}
  \label{fig:image_cleaning_comp}
\end{figure}

Image cleaning is performed in order to reduce the impact of noise on the images.  We use a modified version of the image cleaning algorithm described in \cite{image_cleaning}. An aperture with radius 2.5 pixels is defined, and for each pixel $i$, its weighted luminance is defined as $L_{i} = \sum\limits_{j} l_{j} * w_{j}$, where $j$ is for all pixels, $l_{j}$ is the luminance on it, and $w_{j}$ is the fraction of the $j$-th pixel that lies inside the aperture centered at pixel $i$. Then the background luminance at $i$ is calculated as $B_{i} = T * S_{aperture} * S_{pixel}$, where $T$ is the threshold luminance on pixel that we define, $S_{aperture}$ is the area of aperture, and $S_{pixel}$ is area of one single pixel. If background luminance is larger than weighted luminance on the $i$-th pixel we calculate, the $i$-th pixel is regarded as background and its luminance is set to 0. The effect of image cleaning on the luminance distribution is shown in Figure \ref{fig:image_cleaning_comp}.

\section*{Acknowledgements}

This work is supported by U.S. National Science Foundation Award PHY-2013102.  We are grateful to Ariel Levi Simons and Jeffrey Peacock for valuable collaboration on the DECO project and to Simon Spannagel and Paul Schütze, the developers of Allpix$^2$, for valuable conversations.

\bibliography{DECO_simulation}

\end{document}